\documentclass[prb,twocolumn,showpacs,amsmath,amssymb,superscriptaddress]{revtex4-2}
\usepackage{amsmath,amssymb}
\usepackage[pdftex]{hyperref,graphicx}
\hypersetup{colorlinks = true, urlcolor = blue, linkcolor = blue, citecolor = blue}
\usepackage{physics}
\usepackage{xcolor}
\usepackage{bm}

\newcommand{\comment}[1]{{}}

\newcommand{\cmmnt}[1]{}

\begin{document}
\title{Planckian properties of 2D semiconductor systems}

\author{Seongjin Ahn}
\affiliation{Condensed Matter Theory Center and Joint Quantum Institute, Department of Physics, University of Maryland, College Park, Maryland 20742-4111, USA}
\author{Sankar Das Sarma}
\affiliation{Condensed Matter Theory Center and Joint Quantum Institute, Department of Physics, University of Maryland, College Park, Maryland 20742-4111, USA}

\date{\today}

\begin{abstract}
We describe and discuss the low-temperature resistivity (and the temperature-dependent inelastic scattering rate) of several different doped 2D semiconductor systems from the perspective of the Planckian hypothesis asserting that $\hbar/\tau =k_\mathrm{B}T$ provides a scattering bound, where $\tau$ is the appropriate relaxation time. The regime of transport considered here is well-below the Bloch-Gruneisen regime so that phonon scattering is negligible. The temperature-dependent part of the resistivity is almost linear-in-$T$ down to arbitrarily low temperatures, with the linearity arising from an interplay between screening and disorder, connected with carrier scattering from impurity-induced Friedel oscillations. The temperature dependence disappears if the Coulomb interaction between electrons is suppressed. The temperature coefficient of the resistivity is enhanced at lower densities, enabling a detailed study of the Planckian behavior both as a function of the materials system and carrier density.  
Although the precise Planckian bound never holds, we find somewhat surprisingly that the bound seems to apply approximately  with the scattering rate never exceeding $k_\mathrm{B} T$ by more than an order of magnitude either in the experiment or in the theory. 
In addition, we calculate the temperature-dependent electron-electron inelastic scattering rate by obtaining the temperature-dependent self-energy arising from Coulomb interaction, also finding it to obey the Planckian bound within an order of magnitude at all densities and temperatures.
We introduce the concept of a generalized Planckian bound where $\hbar/\tau$ is bounded by $\alpha k_\mathrm{B} T$ with $\alpha\sim 10$ or so in the super-Planckian regime with the strict Planckian bound of $\alpha=1$ being a nongeneric finetuned situation.

\end{abstract}

\maketitle
\section{Introduction} \label{sec:introduction}
An intriguing empirical observation was made by Bruin, et al., \cite{bruinSimilarityScatteringRates2013} in 2013: The transport scattering rate as extracted from the Drude resistivity formula, $\rho = m/(ne^2\tau)$, where $m$, $n$ and $\tau$ are the carrier effective mass, carrier density, and the transport relaxation time respectively, obeys an approximate bound, $\hbar/\tau = k_\mathrm{B}T$, for many different metals, particularly in the regime where $\rho(T)$ is approximately linear in the temperature (often, such conductors manifesting a linear-in-$T$ resistivity are dubbed `strange metals', particularly if the linearity persists to rather low temperatures). It is implied that $\rho(T)$ should really be the $T$-dependent part of the resistivity with the elastic disorder scattering contribution at $T=0$ subtracted out through careful extrapolation. The terminology `Planckian bound or limit' has stuck to this puzzling empirical phenomenon for historical reasons \cite{hartnollPlanckianDissipationMetals2021a}\cmmnt{[2]} with many experiments \cite{grissonnancheLinearinTemperatureResistivity2021, legrosUniversalTlinearResistivity2019,caoStrangeMetalMagicAngle2020, yuanScalingStrangemetalScattering2022a, huangNonFermiLiquidTransport2020, senStrangeSemimetalDynamics2020, wangUnconventionalFreeCharge2020, ayresIncoherentTransportStrangemetal2021, lizaireTransportSignaturesPseudogap2021, taupinAreHeavyFermion2022}, mostly in 2D strongly correlated systems such as cuprates, claiming the observation of such Planckian bounds on the resistive scattering rates. Exceptions to the Planckian bound on transport have also been pointed out in a few situations \cite{poniatowskiCounterexampleConjecturedPlanckian2021, collignonHeavyNondegenerateElectrons2020}. Such exceptions, where the effective Drude scattering rate is larger than the putative Planckian thermal bound is referred to as the super-Planckian behavior and, by contrast, the situation of the rate being much less than temperature is called sub-Planckian. Particular significance is often attached to the Planckian bound being saturated, i.e., $\hbar/\tau=k_\mathrm{B} T$ (``Planckian metals'') and it is sometimes asserted that the bound is an intrinsic limit on temperature-dependent transport except perhaps for trivial temperature-independent elastic scattering at $T=0$.

The Planckian bound obviously cannot apply to the total resistivity of a metal since all metals have a disorder-induced `residual resistivity' at low temperatures (ignoring any superconducting transition), where the bound must be increasingly violated with the lowering of temperature. The bound must therefore be formally defined by writing:
\begin{equation}
   \rho = \rho_0 + \rho (T)
   \label{eq:total_resistivity}
\end{equation}
where $\rho(T)$ is the temperature-dependent part of the resistivity which vanishes at $T=0$, and $\rho_0$ is, by definition, the disorder-induced residual resistivity at $T=0$. In discussing Planckian properties, it is always implicitly assumed that $\rho(T)$ is being considered in extracting the scattering rate, with $\rho_0$ subtracted out (or equivalently the situation $\rho(T) \gg \rho_0$ applies). In principle, one should worry about the applicability of the Matthiessen's rule in separating out different scattering contributions, particularly in electron systems with low Fermi temperatures $T_\mathrm{F}$ as may happen for certain strongly correlated 2D metals (but not for regular 3D normal metals), but as long as $\rho(T)\gg\rho_0$ applies, this is not a problem. We will always focus on the temperature-dependent part of the resistivity in discussing Planckian properties in the current work even if it is not always explicitly stated everywhere.

An important point, not often emphasized in discussing Planckian transport, but was already discussed in Ref.~\cite{bruinSimilarityScatteringRates2013},  is that regular 3D normal metals may violate the Planckian bound for $T>40K$ or so, where the metallic resistivity is linear-in-$T$ caused by acoustic phonon scattering in the equipartition regime. For example, Pb manifests a linear-in-$T$ room temperature resistivity which, when converted to a transport scattering rate via the Drude formula, gives $\hbar/\tau \sim 9 k_\mathrm{B}T$, reflecting an empirical super-Planckian behavior strongly violating the Planckian bound. By contrast, Al (and many other metals), obeys the Planckian bound at all temperatures.  Basically, all strong-coupling (in the electron-phonon interaction sense) metals with the effective dimensionless electron-phonon coupling strength $\lambda \sim 1$ are trivially super-Planckian since the corresponding transport scattering rate in the linear-$T$ resistivity regime ($T>40K$) is given by 
$(\hbar/\tau) / ( k_\mathrm{B}T) \sim 5$--$10$. 
Therefore, the Planckian bound may often be violated at higher temperatures in metals by the trivial electron-phonon interaction in the quasi-elastic equipartition scattering regime \cite{hwangLinearinResistivityDilute2019}. It has therefore been suggested that any serious discussion of the Planckian bound should also leave out phonon scattering, in addition to leaving out impurity scattering, since high-temperature phonon scattering is quasielastic, and the bound does not apply to any type of elastic scattering (although, as a matter of empirical fact, the bound does apply to phonon scattering limited resistivity in metals, within a factor of $10$). Of course, this makes the whole Planckian analysis something of a theoretical semantic exercise since experimentally all one can measure is the electrical resistivity and convert it into a scattering rate (by also measuring the effective mass and the effective carrier density), and there is no empirical way of ascertaining whether the resistive scattering is or is not elastic/quasielastic. Also, the subtraction of the disorder induced residual resistivity always involves some arbitrariness as it requires an extrapolation to $T=0$.

In fact, resistive scattering is associated with momentum relaxation, and any process, whether elastic or inelastic, leads to a resistivity if it leads to a relaxation of the net momentum. Nevertheless, the whole Planckian lore has taken on considerable significance because of its claimed connection to `strange metals', where a linear-in-$T$ resistivity arises from some unknown electron correlation effects, and somehow persists to low temperatures. In the literature, researchers often conflate the resistive scattering rate with the imaginary part of a single-particle self-energy arising from electron-electron interactions although it is well-known that the two quantities generally have nothing to do with each other as self-energy (resistivity) is associated with the single-particle (two-particle) propagator. In addition, the standard electron-electron scattering rate arising from the imaginary part of the electron self-energy goes as $T^2$ in 3D (or $T^2 \ln{T}$ in 2D), and therefore, any associated resistive scattering would manifest a $T^2$ resistivity at low temperatures (and not a linear-in-$T$ strange, metallic resistivity). 
This $T^2$ versus $T$ issue is sometimes turned around to insist: (1) The linear-in-$T$ resistivity of the so-called strange metals indeed arises from electron-electron interaction behaving in some unknown `strange' manner, and (2) therefore, the physics here must be strange since electron-electron correlation effects produce a linear-in-$T$ resistivity instead of a $T^2$ resistivity. In our opinion this is incorrect logic unless one can show that a reasonable microscopic model leads to correlation-induced linear-in-$T$ resistivity. Often, such a linear-in-$T$ resistivity is somewhat vaguely associated with quantum criticality, but to the best of our knowledge, there is no known physical itinerant electron quantum critical point leading to a linear-in-$T$ resistivity or Planckian scattering.
It should be emphasized that electron-electron scattering can relax momentum only if umklapp or interband Baber scattering is invoked, and connecting an electron self-energy directly with a transport scattering rate is in general incorrect because all momentum dependence is being ignored uncritically in such considerations. The experimental literature is filled with uncritical (and often incorrect) claims of a system being a non-Fermi-liquid simply because it manifests a linear-in-$T$ resistivity at low temperatures, with the unreasonable assumption being that the observed linear-in-$T$ resistivity necessarily implies an imaginary part of self-energy going also as linear in $T$, which would indeed be inconsistent with a Fermi liquid. Such a non-Fermi-liquid claim necessitates at the minimum the observation of the inelastic scattering rate going as $\mathcal{O}(T)$ at arbitrarily low temperatures, not just the resistive scattering rate.

The purpose of the current work is to consider and analyze a well-studied \cite{dassarmaScreeningTransport2D2015,dassarmaMetallicityItsLowtemperature2004,dassarmaLowdensityFinitetemperatureApparent2003,zalaInteractionCorrectionsIntermediate2001} problem, namely the low-temperature density and temperature dependent resistivity of high quality 2D semiconductor systems, from the perspective of the Planckian behavior. In a narrow sense, our work has some superficial similarity with Ref.~\cite{bruinSimilarityScatteringRates2013} where the experimental temperature dependent resistivity of various metals (both 3D and 2D) was analyzed from the Planckian perspective reaching the purely empirical conclusion that most metals obey the Planckian bound. The big qualitative difference between Ref.~\cite{bruinSimilarityScatteringRates2013} and our work is that, in addition to analyzing the existing experimental transport data in 2D semiconductors, we also provide the underlying transport theory which is in approximate agreement with the experimental data, finding to our considerable surprise that the Planckian bound appears to be always obeyed within a factor of $10$. Unlike Ref.~\cite{bruinSimilarityScatteringRates2013}, where all the 3D metallic linear-in-$T$ resistivity most definitively arises from acoustic phonon scattering, our work involves no phonon scattering at all since the experiments (and the associated theory) we consider are all restricted to $<10K$, where phonons are thermally suppressed and, therefore, all phonon scattering contribution to the resistivity is strongly suppressed (the so-called Bloch-Gruneisen regime). 
Phonons play no role in the results we discuss here although the same systems and samples do manifest the expected phonon-induced linear-in-$T$ resistivity at higher temperatures ($>10$--$20$K) \cite{minInterplayPhononImpurity2012}.

The experimentally observed strong approximate linear-in-temperature resistivity at low temperatures in many dilute 2D semiconductor systems arises from an interplay between Coulomb disorder and electron-electron interaction, where the carriers scatter from the momentum-dependent screened disorder, which becomes strongly temperature dependent around $2k_\mathrm{F}$ because of the 2D Fermi surface anomaly \cite{sternCalculatedTemperatureDependence1980,goldTemperatureMaximalConductivity1985, dassarmaTheoryFinitetemperatureScreening1986}. Since at low temperatures the most important resistive scattering is the $2k_\mathrm{F}$ back scattering, the strong metallic temperature  dependence of the 2D resistivity arises from the nonanalytic temperature dependence of the 2D polarizability function at $2k_\mathrm{F}$\texttt{--} increasing temperature weakens screening, leading to larger resistivity with increasing temperature, and this effect in the leading-order is linear in $T/T_\mathrm{F}$, where $T_\mathrm{F}$ is the Fermi temperature. Thus, for dilute systems, where $T_\mathrm{F}$ is relatively low, the linear-in-$T$ temperature correction to the residual resistivity at $T=0$ could be large, extending to arbitrarily low temperatures. Such a linear temperature correction $\rho(T)$ to $\rho_0$ arising entirely (i.e., no phonons) from screened disorder scattering violates the textbook Sommerfeld expansion which asserts that all thermal corrections in a Fermi system must go as $\mathcal{O}((T/T_\mathrm{F})^2)$ because of the thermal broadening of the Fermi distribution. This is an unexpected and counter-intuitive result arising from the fact that 2D systems have nonanalytic interaction corrections associated with Fermi surface anomalies. 
This strong temperature dependence of the 2D resistivity is thus a combined effect of disorder and interaction, and disappears theoretically if one assumes, by ignoring the electron-electron Coulomb interaction, the disorder to be unscreened short-range (or long-range) disorder. The temperature dependence is also suppressed if the disorder is screened by long wavelength Thomas-Fermi screening neglecting the momentum dependence because the 2$k_\mathrm{F}$-anomaly disappears in the long wavelength approximation.
We note that disorder breaks translational invariance allowing interaction to affect momentum relaxation indirectly through the screening of disorder.
The interacting 2D system is still a Fermi liquid with well-defined quasiparticles, but subtle nonanalytic corrections associated with electron-electron interactions give rise to Fermi surface anomalies leading to thermal corrections to all quasiparticle properties violating the $T^2$ law of Sommerfeld expansion \cite{buterakosPresenceAbsenceTwodimensional2021a}.  For the temperature dependent resistivity, the physically appealing way of thinking about the linear-in-$T$ correction to the residual resistivity is that the effective momentum-dependent disorder, including interaction effects (i.e., screened disorder), is strongly temperature dependent although the bare disorder arising from quenched impurities obviously is not. An equivalent statement would be that the electrons resistively scatter from the strongly temperature dependent Friedel oscillations associated with the renormalized impurity potential. Note that Friedel oscillations are characteristics of of finite momentum screening at $2k_\mathrm{F}$, and vanishes in the long-wavelength Thomas-Fermi screening approximation often used theoretically.
Thus, the low-temperature linear-in-$T$ resistivity behavior in 2D semiconductor systems is a combined effect of disorder and interaction, extending all the way to $T=0$, but it implies no violation of the Fermi liquid theory. Perhaps this example should be instructive for other systems where an observed linear-in-$T$ resistivity at lower temperatures is automatically claimed to imply a non-Fermi liquid ground state. Linear-in-$T$ resistivity can indeed arise from indirect effects of electron-electron interactions without affecting the Fermi liquid nature of an electronic system. The effect disappears if the electron-electron interaction is set to zero.

Given that the electron-electron interaction plays the decisive role in producing the Fermi surface anomaly leading to a linear-in-$T$ resistivity in 2D semiconductors arising from screened disorder scattering, with electron-phonon scattering being absent in the physics, studying the Planckian properties of 2D semiconductor transport manifesting linear-in-$T$ resistivity takes on great significance because this is a system where the mechanism underlying $\rho(T)$ is understood and there is a huge amount of experimental data covering many different 2D materials manifesting a strong $T$-dependent low-$T$ resistivity.
We emphasize that unscreened disorder in high quality 2D semiconductors arises from unintentional random quenched charged impurities invariably present in the environment, and theoretically, such unscreened disorder by itself produces the expected $\mathcal{O}((T/T_\mathrm{F})^2)$ at low temperatures as implied by the Sommerfeld expansion of the Fermi function since the noninteracting system has no Fermi surface anomalies. Additionally, such unscreened Coulomb disorder gives rise to an `insulating' resistivity, with $\rho(T)$, in fact, decreasing with increasing $T$ as $\mathcal{O}((T/T_\mathrm{F})^2)$ since the thermal smearing of the Fermi surface decreases the effective scattering momentum in the denominator of the Coulomb potential, effectively enhancing the disorder scattering \cite{sternSelfconsistentTreatmentScreening1985}. Thus, the Fermi surface anomaly induced temperature dependent screening effect has a nontrivial qualitative effect on $\rho(T)$, modifying a negative $\mathcal{O}((T/T_\mathrm{F})^2)$ term into a positive $\mathcal{O}(T/T_\mathrm{F})$ correction. Whether such a linear-in-$T$ resistivity is Planckian or not is indeed an important question. We note that for the doped 2D semiconductors of interest in the current work, the system is essentially a continuum electron liquid with all effects of the periodic lattice (and band structure) subsumed in the effective mass and the effective background dielectric constant\texttt{--} this is the extremely successful effective mass approximation utilized universally in the theories of semiconductor transport. Electron-electron interaction induced umklapp scattering is irrelevant in the transport problem of our interest here since the Brillouin zone filling is a minuscule $10^{-5}$ or so with all the carriers being essentially at the band bottom. Thus, all interaction effects enter the theory indirectly through the Coulomb interaction and the 2D polarizability function controlling the screening of random disorder\texttt{--} direct electron-electron scattering is completely momentum-conserving in our systems of interest and does not contribute to the resistivity.

The second part of our work presented here is independent of the resistivity issue, focusing on the calculation of the electron-electron interaction induced inelastic scattering rate, to be compared with the Planckian hypothesis.
As emphasized above, carrier resistivity is associated with momentum relaxation, and not with any imaginary self-energy arising from interaction-induced electron-electron scattering although the two are often conflated uncritically in the discussion on strange metals and Planckian properties. In some theories, where the momentum dependence is ignored and all scattering is by assumption umklapp scattering, the resistivity is given by the momentum independent self-energy, but such theories are typically uncontrolled in any parameter regime.  For the 2D doped semiconductors, however, the interacting self-energy calculation we carry out in this work within a many body theory for the continuum electron liquid is exact in the high-density or small $r_s$ limit, where $r_s$ is the dimensionless Wigner-Seitz radius going as the inverse square-root of the 2D carrier density, since the many body perturbation expansion is exact in the small-$r_s$ limit for Coulomb interaction. We obtain the imaginary part of the electron self-energy in the leading order infinite ring diagram approximation to calculate the interaction-induced temperature dependent inelastic scattering rate as a function of temperature, finding that the Planckian bound is approximately (within one order of magnitude) valid over the whole temperature regime ranging from $T\ll T_\mathrm{F}$ to $T\gg T_\mathrm{F}$ and in between. This is again a surprising and potentially important result establishing explicitly that the Planckian bound indeed applies (at least approximately and empirically) to the imaginary part of the dynamical self-energy at all temperatures, i.e., to the inelastic single-particle scattering rate arising from electron-electron Coulomb coupling.

This paper thus presents a study of three independent properties of 2D doped semiconductors interconnected only by their relevance to the Planckian hypothesis. The first part (Sec.~\ref{sec:2}) is purely empirical, following the spirit of Bruin et al.~\cite{bruinSimilarityScatteringRates2013}, where we analyze the published low-temperature metallic ($<10K$) experimental resistivity in the context of Planckian. The other two parts (Sec.~\ref{sec:3} and ~\ref{sec:4}) are theoretical with the second part (Sec.~\ref{sec:3}) providing the transport theory for 2D transport which effectively (and approximately) describes the metallic $T$-dependent resistivity discussed in the first part (Sec.~\ref{sec:2}) using the model of carrier scattering from screened Coulomb disorder, both analytically and numerically, through the Boltzmann-RPA effective theories. The third part (Sec.~\ref{sec:4}) describes the theory for the finite-temperature imaginary part of the 2D electron self-energy, studying the inelastic electron-electron scattering rate in the Planckian context.  We emphasize that this third part (i.e., the imaginary 2D self-energy) in our systems has nothing to do with the transport properties discussed in the first two parts since Galilean invariance in our continuum effective mass system ensures that the electron-electron scattering is strictly momentum-conserving  (e.g., no umpklapp)  and does not contribute to the resistivity. Electron interactions enter into transport indirectly through screening in the first two parts of our work since the presence of disorder breaks the translational invariance allowing interaction effects to affect transport indirectly by dressing or renormalizing (i.e., screening in our case) the effective disorder scattering. A well-known related effect of electron-electron interactions affecting transport is the logarithmic correction to the 2D conductivity in the diffusive limit (the so-called Altshuler-Aronov effect) \cite{altshulerZeroBiasAnomaly1979}, and the physics we discuss in Sec.~\ref{sec:3} is basically the ballistic counterpart of this `interaction effect' coming specifically through the 2$k_\mathrm{F}$-screening of Coulomb disorder.

The rest of this paper is organized as follows. In Sec.~\ref{sec:2}, we provide a Planckian transport analysis for several different 2D semiconductor systems, by comparing the extracted scattering rate from the measured resistivity (taken from the existing published experimental literature) to the temperature over a large temperature and density range.  In Sec.~\ref{sec:3}, we provide the transport theory in approximate agreement with the results in Sec.~\ref{sec:2}, by considering carrier scattering from temperature-dependent screened effective disorder arising from random charged impurities, again comparing the theoretical transport scattering rate with temperature in order to test the Planckian hypothesis. In Sec.~\ref{sec:4}, we calculate the finite-temperature inelastic scattering rate arising from electron-electron Coulomb interaction by obtaining the finite-temperature electron self-energy at arbitrary temperature and density, comparing the inelastic scattering rate with temperature from the Planckian hypothesis. 
In Sec.~\ref{sec:5}, we provide some intuitively appealing heuristic dimensional arguments supporting the approximate existence of a Planckian dissipation bound.
We conclude in Sec.~\ref{sec:6} by discussing the implications of our findings in the context of the extensive current debate in the literature on the relevance of the Planckian bound in strange metals.

\section{Planckian analysis of experimental resistivity} \label{sec:2}

Transport properties of doped 2D semiconductor systems are among the most studied \cite{andoElectronicPropertiesTwodimensional1982,spivakColloquiumTransportStrongly2010, dassarmaElectronicTransportTwodimensional2011}\cmmnt{[Ando Fowler Stern RMP 1982]} electronic phenomena in all of physics, going back to 1966 when the two-dimensional nature of interface-confined electrons in Si-SiO$_2$ inversions layers in Si MOSFETs was first demonstrated \cite{fowlerMagnetoOscillatoryConductanceSilicon1966}\cmmnt{[Fowler PRL 1966]}. Many of the seminal experimental discoveries in condensed matter physics over the last 50 years were first reported in various 2D semiconductor systems, including 
IQHE \cite{klitzingNewMethodHighAccuracy1980}\cmmnt{[von Klitzing PRL 1980]}, 
FQHE \cite{tsuiTwoDimensionalMagnetotransportExtreme1982, suenObservationFractionalQuantum1992a}\cmmnt{[Tsui PRL 1982]}, 
even denominator FQHE \cite{willettObservationEvendenominatorQuantum1987}\cmmnt{ [Willett PRL 1987]}, 
strong localization \cite{mottAndersonTransition1975}\cmmnt{[Mott, Pepper]}, 
2D plasmons \cite{allenObservationTwoDimensionalPlasmon1977}\cmmnt{ [Tsui PRL 1975]}, 
weak localization \cite{bishopNonmetallicConductionElectron1980}\cmmnt{[Bishop PRL 1980]}, 
2D Wigner crystals \cite{tsuiTwoDimensionalMagnetotransportExtreme1982, spielmanResonantlyEnhancedTunneling2000}\cmmnt{[Tsui]}, 
excitonic superfluidity \cite{kelloggVanishingHallResistance2004}\cmmnt{[Eisenstein]}, 
2D metal-insulator crossover \cite{dassarmaScreeningTransport2D2015, sarmaSocalledTwoDimensional2005}\cmmnt{[ SDS Hwang Scientific Reports and Solid State Commun]} and many more. There is a huge amount of published experimental resistivity data available in the literature for various 2D semiconductors, both for electrons ($n$) and holes ($p$) as a function of carrier density and temperature. We focus on 6 typical experimental data sets covering 3 different 2D materials systems ($n$-GaAs, $p$-GaAs, $n$-Si), extracting the transport scattering rate from the Drude formula, subtracting out the $T=0$ extrapolated residual resistivity to focus on the purely $T$-dependent part, $\rho(T)$, of the measured resistivity:
\begin{equation}
    \rho(T) = \frac{m}{n e^2 \tau (T)}.
    \label{eq:finite_temperature_resistivity}
\end{equation}
Note that experimentally one measures the full $\rho$ of Eq.~(\ref{eq:total_resistivity}), and $\rho(T) = \rho - \rho_0$, with $\rho_0$ obtained by an extrapolation to $T=0$ as described below. As explained in the Introduction (Sec.~\ref{sec:introduction}), this subtraction takes out the strictly elastic temperature-independent contribution to the resistivity, which is obviously outside the scope of any Planckian consideration. Once $\rho(T)$ is extracted from the data extrapolation and subtraction, we can obtain the scattering time $\tau(T)$ from Eq.~(\ref{eq:finite_temperature_resistivity}) since $n$ and $m$ are experimentally known in our semiconductor systems.

\begin{figure}[!htb]
  \centering
  \includegraphics[width=\linewidth]{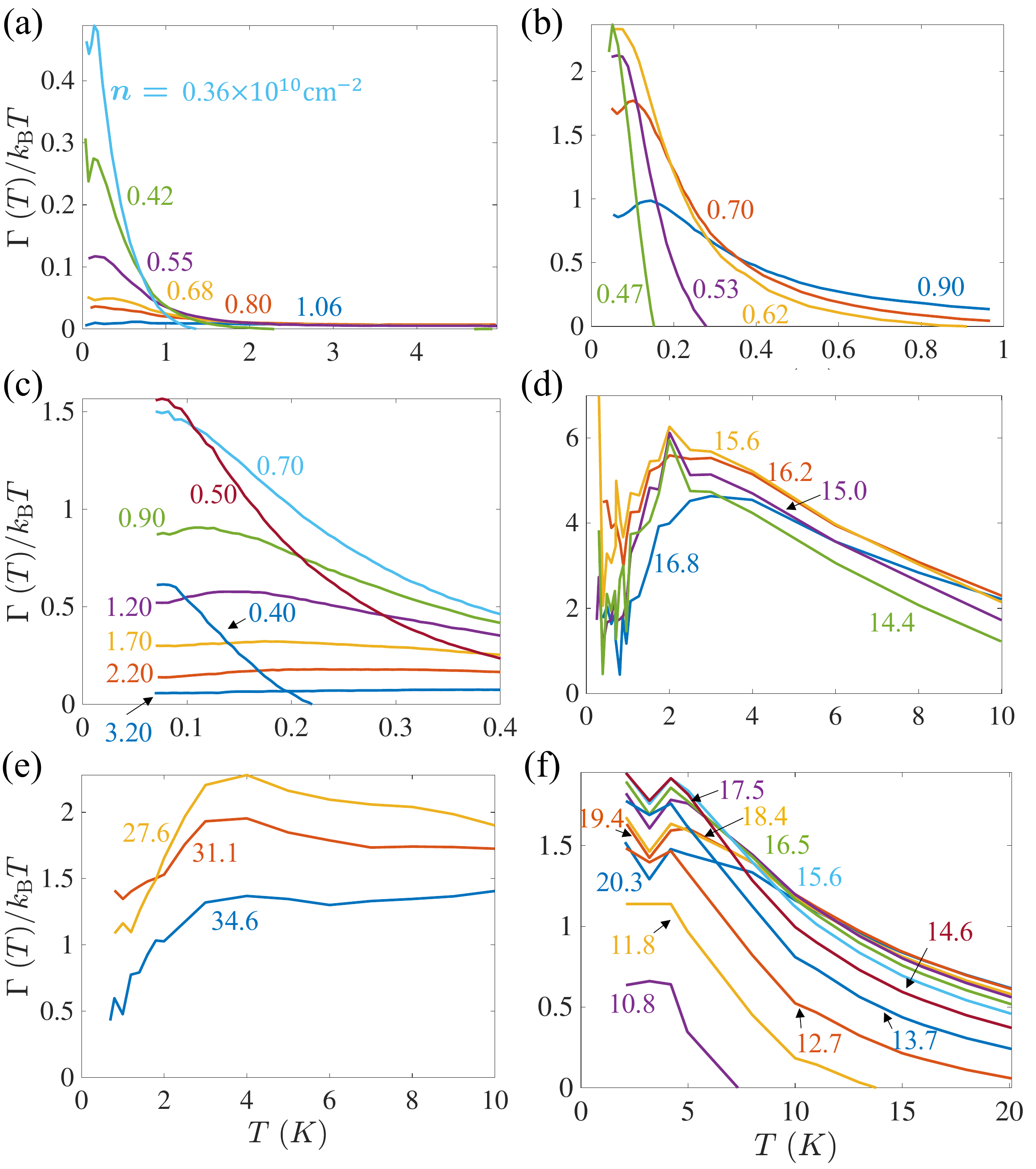}
  \caption{Planckian parameters [defined as $\Gamma(T) /k_\mathrm{B}T$ where $\Gamma(T)=\hbar/\tau(T)$] as a function of $T$, calculated using experimental resitivities of various semiconducting materials [(a) $n$-GaAs \cite{lillyResistivityDilute2D2003a} (b), (c) $p$-GaAs \cite{nohInteractionCorrectionsTwodimensional2003, manfraTransportPercolationLowDensity2007a} and (d)-(f) $n$-Si \cite{tracyObservationPercolationinducedTwodimensional2009a, hwangValleydependentTwodimensionalTransport2013}. Each line with a different color corresponds to a different carrier density $n$ with numbers along the lines representing the corresponding carrier density in the units of $10^{10}\mathrm{cm}^{-2}$. }  
  \label{fig:1}
\end{figure}

In Fig.~\ref{fig:1}, we show the extracted (from the experimental resistivity) dimensionless `Planckian resistivity parameter' $\Gamma/(k_\mathrm{B} T)$, with $\Gamma=\hbar/\tau$ as a function of $T$ for six different 2D samples [Fig.~\ref{fig:1}(a)-(f)] for different carrier densities in each case. The six samples correspond to 3 data sets for $n$-Si, 2 data sets for $p$-GaAs, and 1 data set for $n$-GaAs 2D systems. Since different samples have different disorder and also somewhat different effective thickness of the 2D confinement layers, the results for the resistivity differ from sample to sample even for the same materials (just as the resistivity of different Al samples would differ from sample to sample at low temperatures because of the variations in disorder content).

The important point to note is that in all of these results the dimensionless Planckian parameter (even at its maximum peak value) is always less than $10$, and is often less than $2$. There is considerable noise in the results of Fig.~\ref{fig:1} because of the subtraction of $\rho_0$ through $T=0$ extrapolation and because the original experimental resistivity already has quite a bit of noise in it (and is typically plotted on a log scale, making the extraction of $\tau$ from the published results subject to some inherent errors).

We emphasize a particularly sailent feature of the results in Fig.~\ref{fig:1} with significant relevance to the Planckian debate. As is obvious from Fig.~\ref{fig:1}, there is nothing special about the precise Planckian bound with $\Gamma=k_\mathrm{B} T$, and the experimental $\Gamma/(k_\mathrm{B} T)$ rises above the Planckian bound and then drops below it smoothly as a function of temperature at all densities in all samples with nothing special happening at the Planckian point of $\hbar/\tau=k_\mathrm{B} T$. By fine-tuning and post-selection, one could choose a set of results [see, e.g., $0.9$ density curve in Fig.~\ref{fig:1}(c) and $34.6$ density curve in Fig.~\ref{fig:1}(e)] where the Planckian bound $\hbar/\tau=k_\mathrm{B} T$ holds approximately over a finite $T$-range, but this would reflect purely non-generic confirmation bias since the whole set of results presented in Fig.~\ref{fig:1} for many samples over large ranges of temperature and carrier density clearly establish empirically that there is nothing special about the precise Planckian bound, with $\Gamma/(k_\mathrm{B} T)$ varying above or below unity smoothly\texttt{--}in fact, in Fig.~\ref{fig:1}(a) and (d), the ratio remains always below and above unity respectively. The important point is that the dimensionless Planckian parameter never exceeds $10$, thus there indeed seems to be an approximate empirical thermal bound on the temperature-dependent transport scattering rate.
Although we cannot comment definitely on the Planckian transport analysis of other systems published in the literature, the real possibility of a confirmation bias cannot be ruled out, particularly since the actual precise values of $m/n$ in Eq.~(\ref{eq:finite_temperature_resistivity}) are never quite known in strongly correlated materials with complicated Fermi surfaces (in our systems the Fermi surface is always a circle and the relevant parameters are well-known) and, additionally, the subtraction of residual resistivity is always fraught with some errors. We find that the important physics here is not that there are fine-tuned situations where $\hbar/\tau=k_\mathrm{B} T$ Planckian condition may appear to be satisfied, but the surprising fact that the super-Planckian behavior with $\hbar/\tau>k_\mathrm{B} T$ seems to be bounded within a factor of $5$ above the Planckian bound, but the sub-Planckian behavior with $\hbar/\tau < k_\mathrm{B} T$ persists to arbitrarily low vales of the dimensionless Planckian parameter $\Gamma/k_\mathrm{B} T$. The Planckian bound empirically applying within a factor of $10$ in all the semiconductor transport data we analyzed comes as a real surprise to us.

\begin{figure}[!htb]
  \centering
  \includegraphics[width=\linewidth]{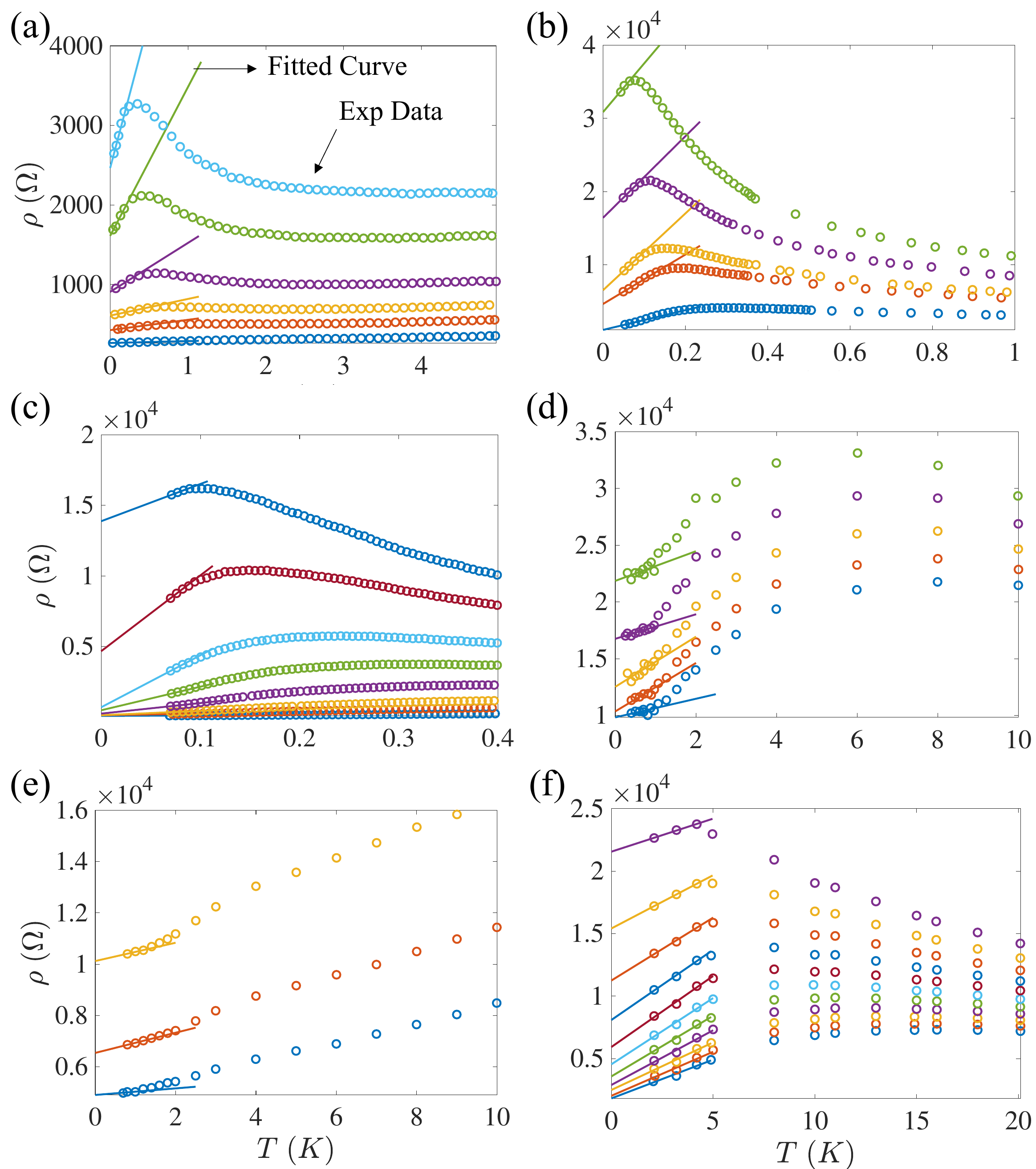}
  \caption{The experimental resistivity data (open circle curves) corresponding to the Planckian results in Fig.~\ref{fig:1} plotted along with the best fitted linear curves (dashed curves) at low temperatures. Each curve corresponds to a different carrier density in the same way as in Fig.~\ref{fig:1}.}  
  \label{fig:2}
\end{figure}

For the sake of completeness, we present in Fig.~\ref{fig:2} the original experimental density- and temperature-dependent resistivity for each sample corresponding to the Planckian results presented in Fig.~\ref{fig:1}, showing also the extrapolation to $\rho_0$ used in our analysis. We emphasize that all our Planckian analysis of the experimental data (Fig.~\ref{fig:1}) is based entirely on the measurements and extrapolations shown in Fig.~\ref{fig:2}. There are numerous other 2D transport results (mostly on Si and GaAs based doped 2D semiconductor systems) in the published literature during 1995-2015, showing very similar results at low temperatures, manifesting strongly $T$-dependent $\rho(T)$ at low temperatures ($<10K$) without any direct electron-phonon scattering seemingly contributing to the metallic resistivity. All of these 2D systems are relatively dilute, i.e., low-density systems with rather low Fermi temperatures ($5K-50K$) and are high-quality, i.e., $\rho_0$ in the metallic regime is typically well below $h/2e^2 \sim 10^4$ohms. The systems undergo a relatively sharp density-induced effective metal-insulator crossover at a strongly system-dependent `critical' density, but our interest is entirely focused on the effective low-temperature metallic phase at densities above this critical density, and we have nothing to say in the current work about the 2D metal-insulator transition itself or the insulating phase for densities below the critical density. Planckian physics applies only to the metallic situation.

In discussing Figs.~\ref{fig:1} and \ref{fig:2}, we start by describing the generic features of the measured resistivity shown in Fig.~\ref{fig:2}: (1) $\rho$ first increases with $T$, rising rather rapidly and approximately linearly at low temperatures, reaching a peak value, which, depending on the system and the carrier density could be 10$\%$ to 100$\%$ larger than $\rho_0$ at $T=0$ within a small overall temperature increase of a few Kelvins; (2) the relative value of the temperature dependence $\rho/\rho_0$ or $\rho(T) / \rho_0$ depends strongly on the carrier density in the same sample, with the temperature dependence increasing with decreasing density; (3) at high density, the temperature dependence, while being linear, is very weak; (4) $\rho(T)$ decreases with increasing $T$ beyond the density- and sample-dependent peak, but at higher densities, where the overall $T$-dependence is generally weak, no such peak in $\rho$ appears; (5) the slow decrease of the high-temperature resistivity with increasing $T$ at lower densities apparent in some of the results of Fig.~\ref{fig:2} is neither a `resistivity saturation' nor `an insulating phase' – it is simply the temperature dependence of a classical metallic resistivity, where increasing $T$ beyond the quantum-classical crossover (occurring at low temperatures for low carrier densities) leads to a slowly decreasing $\rho(T)$ with increasing $T$ \cite{dassarmaLowdensityFinitetemperatureApparent2003}. We note that the temperature scale for the $T$-dependent resistivity in Fig.~\ref{fig:2} is determined by $T/T_\mathrm{F}$ with the Fermi temperature $T_\mathrm{F}$ being proportional to the carrier density, and thus the temperature dependence weakens at higher densities. We emphasize that at lower carrier densities, manifesting the quantum-classical crossover, the typical Fermi temperature in these samples is of $\mathcal{O}(10K)$.

The consequences are that the dimensionless Planckian parameter $\Gamma/(k_\mathrm{B} T)$ plotted in Fig.~\ref{fig:1} as a function of $T$ for various densities typically show a maxima between 0.1K and 5K depending on the sample and the carrier density, but the largest peak value [Fig.~\ref{fig:1}(d)] is only $\sim 6$ whereas most of the peak values are around unity (or below). Thus, one inevitable empirical conclusion based on Fig.~\ref{fig:1} is that the extracted scattering rate is bounded from above by $k_\mathrm{B}T$ within a factor of $2-6$. The largest value of the dimensionless Planckian happens mostly at some intermediate temperatures of $\mathcal{O}(1K)$ for Si-based samples [Figs.~\ref{fig:1}(d)-(f)], whereas for GaAs samples [Figs.~\ref{fig:1}(a)-(c)] the peak is around $\mathcal{O}(0.1K)$. At higher $T$ values, the dimensionless Planckian invariably decreases to strongly sub-Planckian behavior, becoming much less than unity.  We emphasize that the same sample may manifest super-Planckian ($\Gamma> k_\mathrm{B} T$), Planckian ($\Gamma\sim k_\mathrm{B} T$), and sub-Planckian ($\Gamma < k_\mathrm{B}T$) behavior at different densities and temperatures, clearly establishing that the idea of a strange Planckian metallicity with $\Gamma \sim k_\mathrm{B} T$ is not a precise well-defined parameter-independent concept.

What we find to be the most interesting empirical finding in Fig.~\ref{fig:1} is that the dimensionless Planckian parameter in doped 2D semiconductors manifests sub-Planckian ($<1$) or Planckian ($\sim 1$) or super-Planckian ($>1$) behavior in the same sample as the carrier density and temperature are varied. The highest value of the dimensionless Planckian parameter is mostly achieved at the lowest carrier density (where also the $T$-dependence of the measured resistivity in Fig.~\ref{fig:2} is the strongest), and even this highest value typically is of the order of only $2$--$6$, never exceeding the Planckian limit of unity by more than an magnitude.

In the next section, we discuss a Boltzmann transport theory for the empirical results presented in the current section, based on the carriers being resistively scattered by temperature- and momentum-dependent screened Coulomb disorder (i.e., $T$-dependent Friedel oscillations).

\section{Transport Theory} \label{sec:3}

\begin{figure}[!htb]
  \centering
  \includegraphics[width=\linewidth]{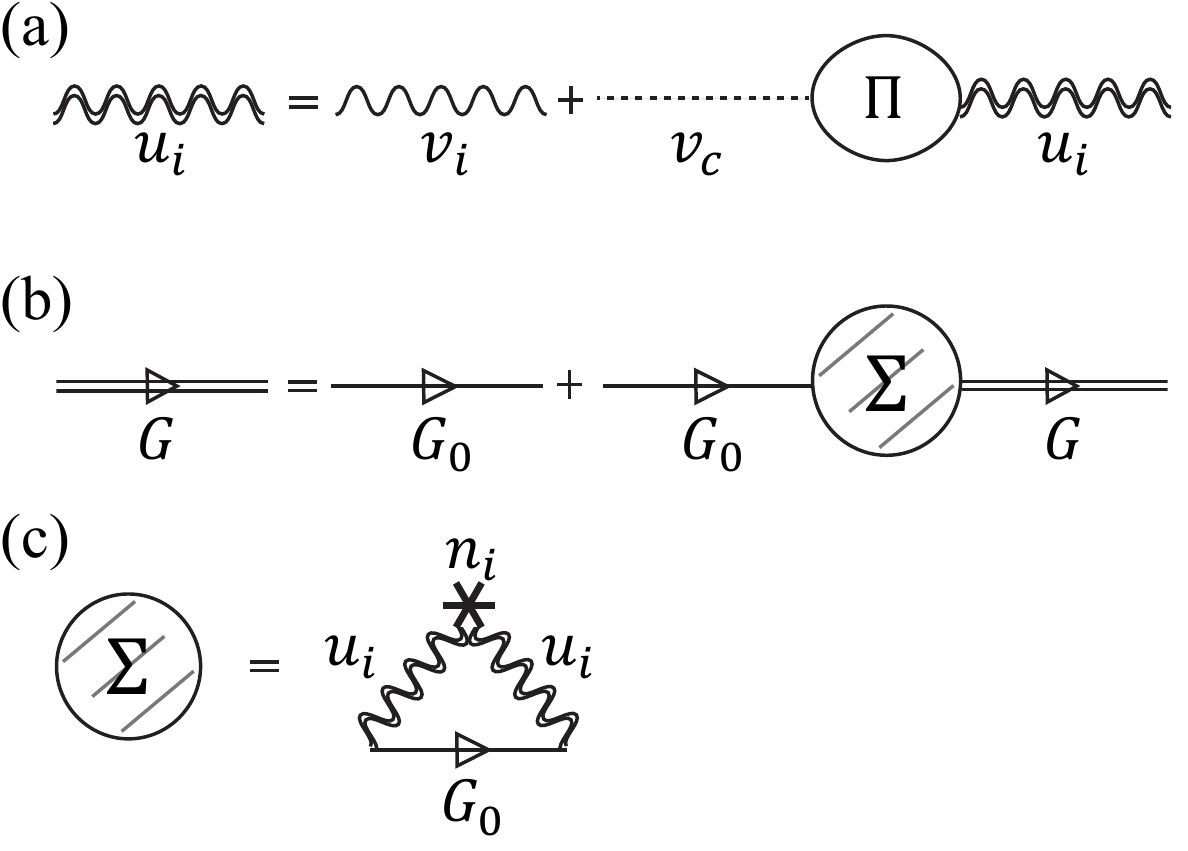}
  \caption{Feynman diagrams for (a) the screened electron-impurity interaction ($u_i$), (b) the single particle impurity Green's function ($G$), and (c) the impurity self-energy ($\Sigma$) within the leading order Born approximation. $n_i$, $G_0$, $v_c$ and $\Pi$ represent the impurity density, the bare Green's function, the bare Coulomb interaction, and the bare polarizability, respectively. Here we suppress the momentum label for visual clarity.  }  
  \label{fig:transport_diagram_figure}
\end{figure}
We consider resistive scattering of 2D carriers with a standard parabolic energy dispersion, with a 2D density of $n$, described by an effective mass $m$ and a background lattice dielectric constant $\kappa$, from random quenched charged impurities of effective 2D concentration $n_i$. The Coulomb coupling between the carriers and the impurities is screened by the momentum-dependent dielectric screening function of the carriers themselves, $\epsilon(q)$, where the density ($n$) and the temperature ($T$) dependence of the screening function is suppressed for notational convenience. We treat the finite-temperature and finite-momentum screening in the mean field random phase approximation(RPA), which expresses the dielectric screening in terms of the electron-electron long-range Coulomb interaction and the 2D noninteracting finite-momentum and finite-temperature polarizability function, $\Pi(q)$ (This RPA screening theory is essentially the leading-order theory in 1/$N$ expansion assuming $N$ fermionic flavors, as used extensively in quantum field theories). Given the screened electron-impurity interaction, we calculate the resistivity using the leading-order Boltzmann transport theory treating the screened disorder scattering in the Born approximation. Figure~\ref{fig:transport_diagram_figure} provides the schematic Feynman diagrams for the equivalent screened disorder approximation for the single-particle Green’s function, but of course we obtain the full transport coefficient including the appropriate vertex correction which is automatically guaranteed by the Boltzmann transport theory. 
Since the interaction between the carriers and the random impurities is finite ranged here (`screened Coulomb disorder'), the vertex corrections are quantitatively important for doped semiconductors unlike in normal 3D metals where the electron-impurity scattering is always $s$-wave with vertex corrections being unimportant for the resistivity.

The basic quantity entering the Boltzmann resistivity formula is the thermally-averaged scattering rate $\tau$ entering Eq.~(\ref{eq:total_resistivity}). The main calculation is obtaining the scattering rate, $1/\tau$, at finite temperatures as a function of carrier density and temperature (in the leading order theory the scattering rate is proportional to the impurity density $n_i$, simply providing an overall resistivity scale). All the equations going into our transport theory are shown below.
The thermal average of the scattering time is given by 
\begin{equation}
    \tau=\frac{  \int d\varepsilon_\mathrm{\bm k}  \tau(\mathrm{\varepsilon_\mathrm{\bm k}}) \varepsilon_\mathrm{\bm k}\left(-\frac{ \partial f(\varepsilon_\mathrm{\bm k})}{\partial\varepsilon_\mathrm{\bm k}}\right)  }
    {\int d \varepsilon_\mathrm{\bm k} \varepsilon_\mathrm{\bm k} \left(-\frac{ \partial f(\varepsilon_\mathrm{\bm k})}{\partial\varepsilon_\mathrm{\bm k}}\right) },
    \label{eq:tau_finite_T}
\end{equation} 
where $\varepsilon_\mathrm{\bm k}=\hbar^2k^2/2m$ is the energy dispersion with $m$ denoting the effective mass, $f(\varepsilon)=1/[1+\exp(\varepsilon-\mu(T))/k_\mathrm{B}T]$ is the Fermi-Dirac distribution function, $\mu(T)$ is the chemical potential with the Fermi energy $E_\mathrm{F} =\mu(T=0)$, and $\tau({\varepsilon_\mathrm{\bm k}})$ is the zero-temperature scattering time, which, in the leading-order Boltzmann transport theory (Fig.~\ref{fig:transport_diagram_figure}), is given by
\begin{equation} 
    \frac{1}{\tau({\varepsilon_\mathrm{\bm k}})}=\frac{2\pi n_i}{\hbar} 
    \sum_\mathrm{\bm k'}
    \left|u_\mathrm{i}(\bm k - \bm k')\right|^2
    (1-\cos{ \theta})\delta(\epsilon_\mathrm{\bm k}-\epsilon_\mathrm{\bm k'}).
    \label{eq:tau_zero_T}
\end{equation} 
Here $u_\mathrm{i}(\bm q)=v_i(q)/\epsilon(q,T)$ is the screened electron-impurity Coulomb interaction with $v_i(q)=2\pi e^2/\kappa q$ and $\epsilon(q,T)=1-v_c(q)\Pi(q,T)$ is the RPA screening function, where $v_c(q)=2\pi e^2/\kappa q$ is the electron-electron Coulomb interaction and $\Pi(q,T)$ is the finite temperature 2D polarizability, which can be obtained by using the zero temperature polarizability:
\begin{equation}
    \Pi(q,T)=\int_\mathrm{0}^{\infty} d\epsilon \frac{\Pi(q)|_{\epsilon_\mathrm{F}=\epsilon}}{4k_\mathrm{B}T\cosh^2{\frac{\epsilon-\mu(T)}{2k_\mathrm{B}T}}},
    \label{eq:pol_finite_T}
\end{equation}
where 
 \begin{equation}
    \Pi(q)=
		\frac{gm}{2\pi\hbar^2}\left[1 - \Theta(q-2k_\mathrm{F})\frac{\sqrt{q^2- 4k^2_\mathrm{F} }}{q}\right],
    \label{eq:pol_zero_T}
\end{equation}
is the zero temperature static 2D polarizability. Here $E_\mathrm{F}(k_\mathrm{F})$ is the Fermi energy(wavevector), $g$ denotes the valley and spin degeneracy and $\Theta(x)$ is the Heaviside step function. 
Finite temperature smoothens the $q=2k_\mathrm{F}$ kink in the 2D polarizability algebraically even for $T\ll T_\mathrm{F}$, thus contributing a linear-in-$T$ nonanalytic correction to the resistivity since $2k_\mathrm{F}$ back-scattering is the dominant scattering contributing to the resistivity\texttt{--} this 2D-specific thermal effect disappears if the momentum-independent long-wavelength Thomas-Fermi approximation is used.  This linear-in-$T$ term is not present in the Sommerfeld expansion or in 3D electrons scattering from screened disorder\texttt{--}the effect is intrinsic to 2D screening properties.

\begin{figure}[!htb]
  \centering
  \includegraphics[width=\linewidth]{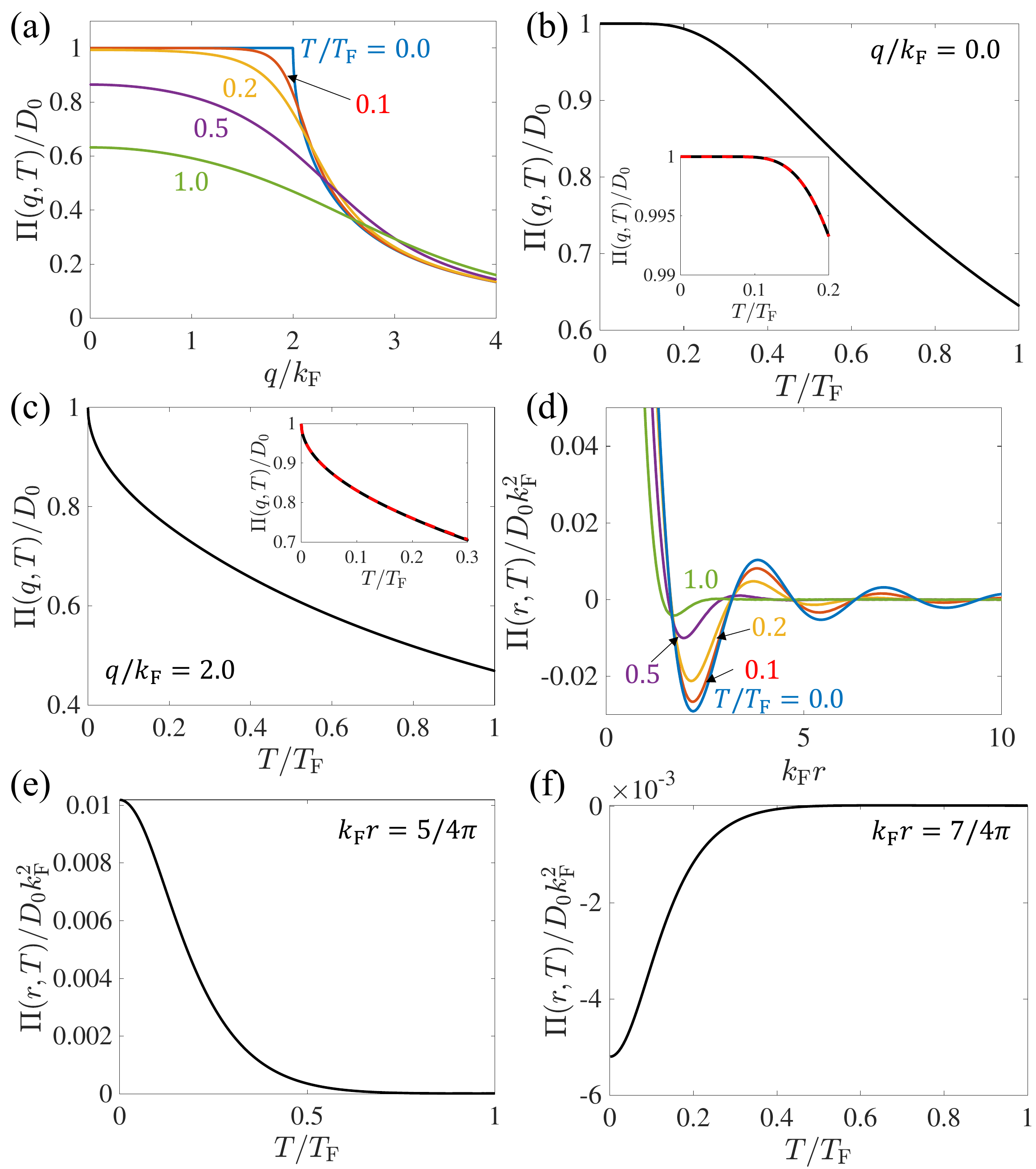}
  \caption{(a) Finite temperature 2D polarizability as a function of (a) wavevector for various temperatures $T/T_\mathrm{F}=0.0$, $0.1$, $0.2$, $0.5$, $1.0$, and as a function of (b), (c) temperature at a fixed wavevector of (b) $q/k_\mathrm{F}=0.0$ and (c) $q/k_\mathrm{F}=2.0$. The insets in (b) and (c) show the zoom-in of the low temperature region, comparing the numerical results (black solid lines) with the known analytical asymptotic forms (red dashed lines) given by $\Pi(q=0,T)/D_0=1-\exp(-T_\mathrm{F}/T)$ and $\Pi(q=2k_\mathrm{F},T)/D_0=1-\sqrt{\pi/4}(1-\sqrt{2})\zeta(1/2)\sqrt{T/T_\mathrm{F}}$. (d)-(f) Finite temperature 2D polarizability in real space as a function of (d) distance $r$ for various temperatures $T/T_\mathrm{F}=0.0$, $0.1$, $0.2$, $0.5$, $1.0$, and as a function of (e), (f) temperature at a fixed distance of (e) $k_\mathrm{F}r=5/4\pi$ and (f) $k_\mathrm{F}r=7/4\pi$, showing a strong temperature dependence in the low temperature regime.}  
  \label{fig:4}
\end{figure}

This is illustrated in our Fig.~\ref{fig:4}, where we plot the calculated finite-temperature 2D polarizability (i.e., the `bubble' in Fig.~\ref{fig:transport_diagram_figure}) as a function of momentum for different values of $T/T_\mathrm{F}$, clearly showing that the thermal smearing of the $2k_\mathrm{F}$-kink is $\mathrm{O}(T^{1/2})$ whereas at long wavelength (i.e., zero momentum), defining the Thomas-Fermi screening, the thermal smearing is exponentially weak, clearly showing that a long wavelength screening approximation, the standard approximation in semiconductor physics, would completely miss the strong temperature dependence observed experimentally.  We also show the  2D polarizability in the real space, bringing out the strong temperature dependence of the Friedel oscillations in screening associated with the strong thermal smearing of the 2$k_\mathrm{F}$-kink in the momentum space.  These strongly temperature dependent Friedel oscillations scatter carriers, leading to the strong temperature dependence of the resistivity arising from the  strong temperature dependence of the effective screened disorder.  the effect is completely lost in any long wavelength screening approximation, but $2k_\mathrm{F}$-scattering is the most important resistive scattering process, and hence the temperature dependence of $2k_\mathrm{F}$ screening is crucial.

It turns out that the above equations allow for a low-temperature ($T\ll T_\mathrm{F}$) and high-temperature ($T\gg T_\mathrm{F}$) analytical expressions for the 2D resistivity by appropriately expanding the finite-temperature screening function.  Note that the analytical expansion is much more sophisticated than the simple Sommerfeld expansion, which would always give $\mathcal{O}(T^2)$ correction at low temperatures arising from the expansion of the Fermi functions.  The thermal expansion involves careful consideration of the nonanalyticity in the 2D polarizability function at $q=2k_\mathrm{F}$, because of the strong $2k_\mathrm{F}$ kink as a function of momentum in the 2D polarizability arising from the Heaviside step function in Eq.~(\ref{eq:pol_zero_T}). The analytical results are:
\begin{equation}
    \begin{aligned}
        \rho(T \ll T_\mathrm{F}) &\approx \rho_0\left[\frac{2x}{1+x}\left(\frac{T}{T_\mathrm{F}}\right) +
                        \frac{2.646x}{(1+x)^2} \left(\frac{T}{T_\mathrm{F}}\right)^{3/2} \right] \\
        \rho(T \gg T_\mathrm{F}) &\approx  \rho_1 \left(\frac{T_\mathrm{F}}{T}\right) \left[1-\frac{3\sqrt x}{4}\left(\frac{T_\mathrm{F}}{T}\right)^{3/2} \right] 
    \end{aligned}
    \label{eq:resistivity_asymptotic_low_and_large_T}
\end{equation}
where $\rho_0=\rho(T=0)$, $\rho_1=(h/e^2) (n_i/n) (2\pi x^2/g^2)$, and $x=q_\mathrm{TF}/2k_\mathrm{F}$, with $q_\mathrm{TF}= 2 m e^2/\kappa\hbar^2$ as the 2D Thomas-Fermi screening wavenumber. The key dimensionless parameters controlling the temperature dependence of resistivity are $x=q_\mathrm{TF}/2k_\mathrm{F}$ and $T/T_\mathrm{F}$\texttt{--} for low(high) density, both are large(small) since $T_\mathrm{F}$ and $k_\mathrm{F}$ are proportional to $n$ and $n^{1/2}$, respectively, (and $q_\mathrm{TF}$ is independent of density in 2D), leading to strong(weak) temperature dependence.


Before presenting our full numerical transport results for various 2D semiconductor systems (without any expansion in $T/T_\mathrm{F}$), we first provide an analytical Planckian analysis of the low-temperature results establishing that the analytical result, rather amazingly, is consistent, within a factor of 4, with the Planckian bound conjecture. Using Eq.~(\ref{eq:total_resistivity}) and the leading linear-order term of Eq.~(\ref{eq:resistivity_asymptotic_low_and_large_T}), we write:
\begin{equation}
    \rho(T) = \rho – \rho_0 = \rho_0 \frac{2x}{1+x}\frac{T}{T_\mathrm{F}}.
    \label{eq:resistivity_asymptotic_low_T}
\end{equation}
We note that $\rho(T)$ is proportional to $\rho_0$, thus increasing linearly with disorder $n_i$ since $\rho_0$ is proportional to $n_i$ in the leading-order transport theory. Since we are interested in the metallic transport property and in the maximum possible value of metallic $\rho(T)$ in order to test the Planckian conjecture, we use the Ioffe-Regel-Mott (IRM) criterion choosing the maximum possible metallic value of $\rho_ 0$ to be the 2D IRM limit of $h/e^2$ when the mean free path equals $1/k_\mathrm{F}$. Putting $\rho_0 = h/e^2$, we automatically obtain the highest possible value of the dimensionless Planckian parameter $(\hbar/\tau)/(k_\mathrm{B} T)$ since the highest $\rho_0$ implies the highest $\rho(T)$ which, by turn, then implies the highest $\hbar/\tau$.  Putting $\rho_0 =h/e^2$ and denoting the corresponding finite-temperature scattering time as $\tau_\mathrm{min}$ (implying that this is the maximum allowed scattering rate $1/\tau_\mathrm{min}$ which is consistent with a 2D metal) and doing simple algebra, we obtain:

\begin{equation}
    \hbar/\tau_\mathrm{min} = 4k_\mathrm{B} T    
\end{equation}

Thus, the maximum possible scattering rate in our theory is limited from above by $4k_\mathrm{B} T$, indicating that the dimensionless Planckian cannot be much larger than $k_\mathrm{B} T$.  If we take into account the fact that our leading order Boltzmann theory itself most likely breaks down substantially below the strong scattering IRM limit is reached, we conclude that:

\begin{equation}\label{eq:equalityeq10}
    \hbar/\tau(T) < 4 k_\mathrm{B} T,    
\end{equation}
which may be construed as the modified Planckian hypothesis for the transport problem under consideration here, i.e., the temperature-dependent 2D resistivity limited by screened Coulomb disorder scattering.  We emphasize that although we use the IRM criterion on $\tau_0$ to obtain the basic inequality on the scattering rate, the limit imposed on the corresponding $T=0$ scattering rate is simply that the Fermi surface is well-defined even in the presence of disorder scattering. Using the IRM limit on $\rho_0$ itself we get the following inequality for the $T=0$ scattering rate $\tau_0$:
\begin{equation}\label{eq:IRM}
    \hbar/2\tau_0 < E_\mathrm{F}
\end{equation} 
which is nothing other than the IRM criterion defining coherent quasiparticles, i.e., with disorder-induced $T=0$ level broadening ($\hbar/2\tau_0$) being less than the Fermi energy. Thus, the generic assumption of coherent quasiparticle transport logically leads to a Planckian bound (within a factor of $4$) on our theoretical transport scattering rate! This shows that the often-claimed  uncritical assertion that the saturation of the Planckian bound implies incoherent non-quasiparticle and non-Fermi liquid transport cannot be generically correct since it fails for our system where the bound is violated by a factor of $4$ at the limit of coherent transport. 
Of course, one could argue that a factor of 4 is not significant for IRM-type dimensional arguments \cite{bruinSimilarityScatteringRates2013}. 

We mention that the correction to the Planckian inequality of Eq.~(\ref{eq:equalityeq10}) arising from the next-to-the-leading-order term of $\mathcal{O}((T/T_\mathrm{F})^{3/2})$ in Eq.~(\ref{eq:resistivity_asymptotic_low_and_large_T}) is rather small for $T<T_\mathrm{F}$ since it is bounded by $1.323/(1 + x)$, and $x\gg1$ for any strong $T$-dependence to manifest itself anyway. This provides an analytical explanation for why the Planckian bound applies to our numerical results within a factor $5$ agreeing with the corresponding experimental results in Fig.~\ref{fig:1} where the bound is always satisfied within a factor of $6$. These theoretical arguments obviously apply only to the 2D doped semiconductor Planckian behavior because it is not known if similar consideration would apply to other Planckian metals studied in the literature.

We find this analytical derivation of an effective Planckian conjecture for our transport phenomenon to be a rather unexpected result since the IRM criterion is a $T=0$ constraint for metallic transport which has nothing to do with temperature, but nevertheless the finite-temperature scattering rate is absolutely constrained from above by $4k_\mathrm{B} T$ based on the IRM constraint imposed on the $T=0$ resistivity. One may be concerned that this upper bound argument is based on the low temperature analytic result [Eq.~(\ref{eq:resistivity_asymptotic_low_T})], and thus it cannot be ruled out that the constraint is not valid at higher temperatures with $\hbar/\tau>4k_\mathrm{B}T$ at $T>T_\mathrm{F}$. In the subsequent discussions below, however, we show numerically (without assuming $T\ll T_\mathrm{F}$) that at all temperatures the Planckian parameter is of the order of only $2$--$4$ in agreement with our modified Planckian hypothesis discussed above by carefully analyzing the high temperature analytical results and presenting the full numerical results.

We mention one other aspect (see the second part of Eq.~(\ref{eq:resistivity_asymptotic_low_and_large_T}) for $T\gg T_\mathrm{F}$) of our analytical results in approximate agreement with experiments on the high-temperature side, where $\rho(T)$ decreases with increasing $T$ in approximately $1/T$ manner in accordance with the experimental finding presented in Sec.~\ref{sec:2}. Given that $\rho(T)$ increases linearly in $T$ for $T\ll T_\mathrm{F}$ and decreases linearly for $T\gg T_\mathrm{F}$ means that there is a quantum-to-classical crossover in the 2D transport around $T \sim T_\mathrm{F}$, where $\rho(T)$ should have a local maxima. This maximum value of $\rho(T)$ is the most important regime for the validity or not of the Planckian hypothesis. Actual numerics shows that this resistivity maximum occurs roughly around $T_\mathrm{F}/3$ with $\rho(T)$ increasing (decreasing) with $T$ below (above).

\begin{figure}[!htb]
  \centering
  \includegraphics[width=\linewidth]{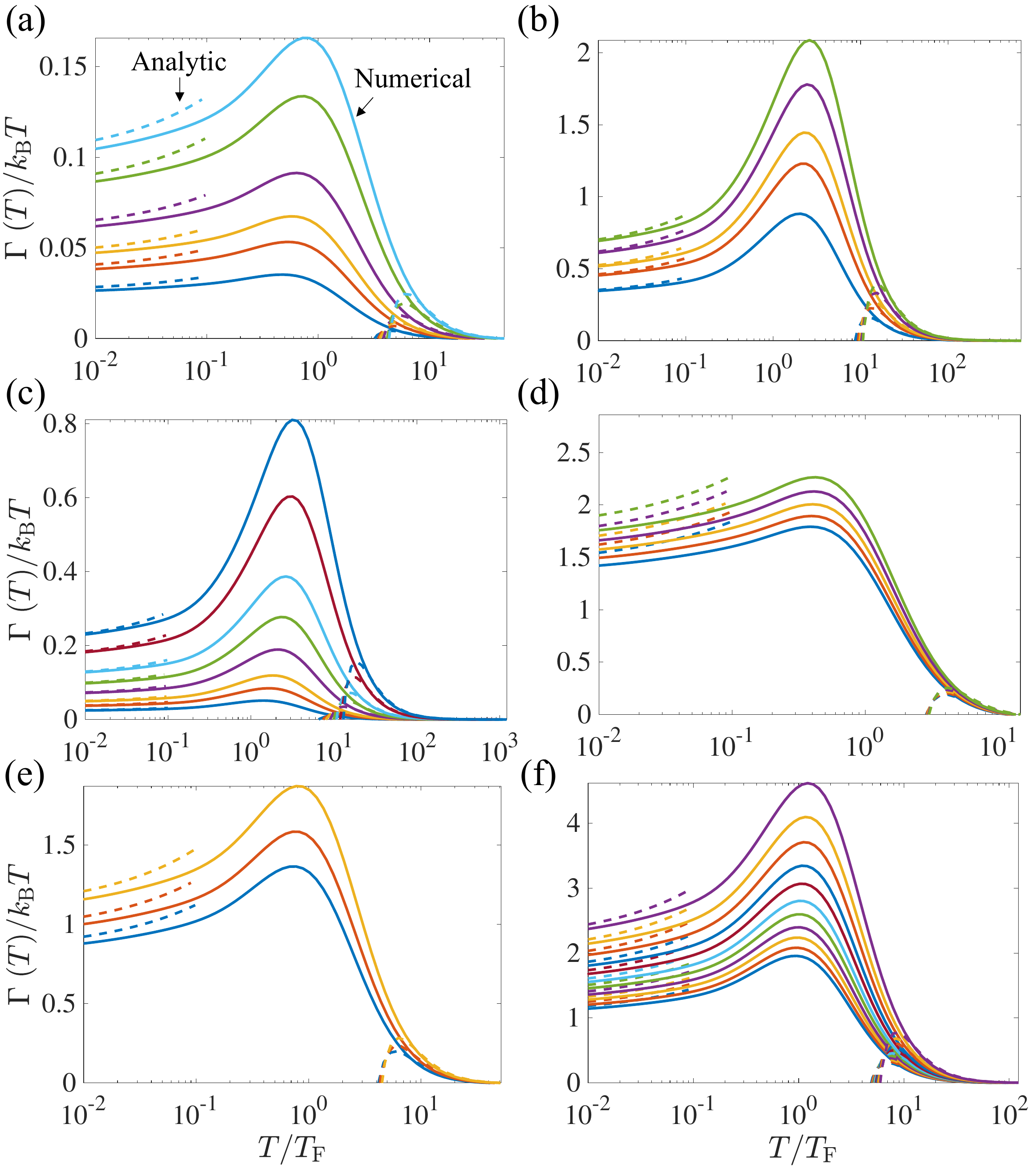}
  \caption{Planckian parameter numerically calculated as a function of $T$ using the Boltzmann transport theory and the material parameters corresponding to the experimental systems in Fig.~\ref{fig:1}/\ref{fig:2}(a)-(f). The dashed curves represent the low- and high-$T$ asymptotic results [Eq.~(\ref{eq:resistivity_asymptotic_low_and_large_T})], which agree well with the full numerical results (solid curves). Each curve corresponds to a different carrier density in the same as way in Fig.~\ref{fig:1}. }  
  \label{fig:5}
\end{figure}

In Fig.~\ref{fig:5}(a)–(f) we present (by solving Eqs.~(\ref{eq:tau_finite_T}-\ref{eq:pol_zero_T}) the full numerically calculated dimensionless Planckian parameter as a function of temperature ($T$) and density ($n$) corresponding to the systems shown in Figs.~\ref{fig:1} and \ref{fig:2} [i.e., using the theoretical materials parameters corresponding to the experimental systems in Figs.~\ref{fig:1}/\ref{fig:2}(a)-(f)]. We also show in each panel of Fig.~\ref{fig:5} the results corresponding to the asymptotic low-$T$ (high-$T$) analytical theory which agree with the full numerical results, deviating from the analytical theory at higher (lower) $T(n)$. Our theory includes the realistic (but nonessential) details of each experimental sample, such as the appropriate valley degeneracy and the quasi-2D width of each 2D system (which typically suppresses the effective interaction through a form factor arising from the confinement wavefunctions) and we assume, to keep the number of parameters a minimum, that the charged impurities are randomly distributed in the 2D layer with an effective 2D density of $n_i$ (relaxing this approximation leads to a better quantitative agreement with the experiment at cost of adding more unknown parameters, which is unnecessary in the context of the current work.) We fix each value of $n_i$ (which only defines the overall scale, not the $T$- and $n$-dependence) by obtaining the best fit with the experiment at the lowest temperature. Thus the random impurity density, $n_\mathrm{i}$, defining the overall resistivity scale, but not its temperature dependence at all, is the only unknown parameter of our model.


We summarize the salient features of our theoretical results presented in Fig.~\ref{fig:5}: (1) The dimensionless Planckian parameter mostly satisfies the Planckian conjecture, within a factor of $10$, being of $\mathcal{O}(1-10)$ or less quite generally; (2) when the parameter exceeds unity, it is only by a factor of $2$--$4$, never by more than an order of magnitude; (3) the theoretical results agree generically qualitatively, and sometimes semi-quantitatively, with the experimental results of Fig.~\ref{fig:1} (this agreement can be made quantitative by using an impurity distribution in the quasi-2D confinement direction, thus adding more parameters to the model in addition to $n_\mathrm{i}$); (4) in general, the Planckian parameter is the largest at the lowest densities [and at intermediate temperatures $\sim\mathcal{O}(T_\mathrm{F}/3)$] for all samples with the behavior being sub-Planckian at higher densities and lower temperatures (again in complete agreement with the experimental data); (5) similar to the experimental findings, the theoretical $\rho (T)$ is linear only for $T\ll T_\mathrm{F}$, deviating from linearity at higher temperatures, but the linearity within the screening theory persists all the way to $T=0$.

We mention in this context that our leading order Boltzmann theory, while being consistent qualitatively with the experimental results everywhere, becomes less valid at lower densities since the theory is exact in the $n\gg n_i$, regime and fails completely for $n<n_i$. The results shown in Figs.~\ref{fig:5} obey the $n\gg n_i$ criterion necessary for the applicability of our Boltzmann theory, and the fitted values of $n_i$ are consistent with the specific materials considered in each case.

\section{Self-energy and INELASTIC SCATTERING RATE} \label{sec:4}
\begin{figure}[!htb]
  \centering
  \includegraphics[width=\linewidth]{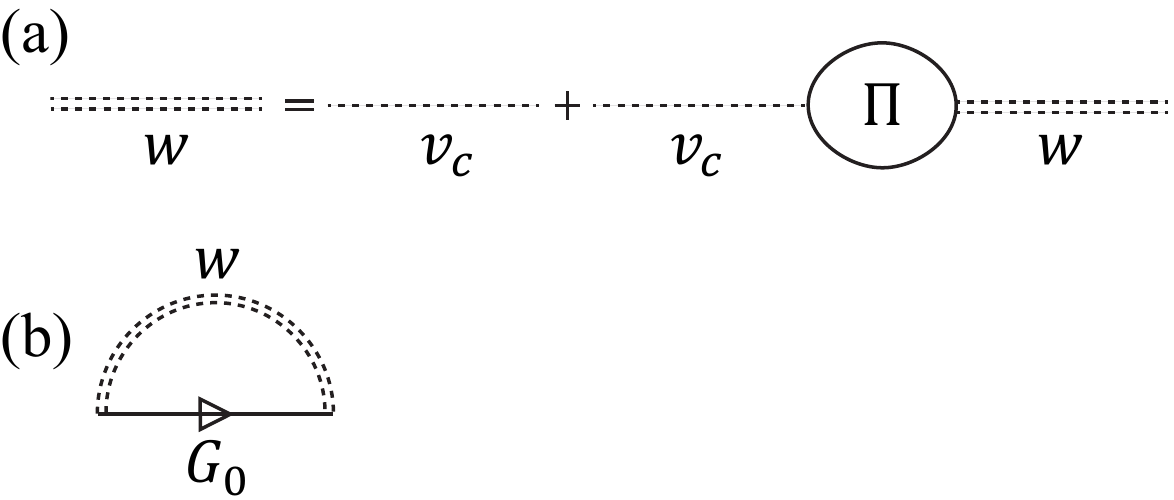}
  \caption{Feynman diagrams for (a) the screened Coulomb interaction($w$), and (b) the RPA self-energy. The notations are the same as in Fig.~\ref{fig:transport_diagram_figure}. }  
  \label{fig:self_energy_diagram_figure}
\end{figure}

\begin{figure*}[!htb]
  \centering
  \includegraphics[width=\linewidth]{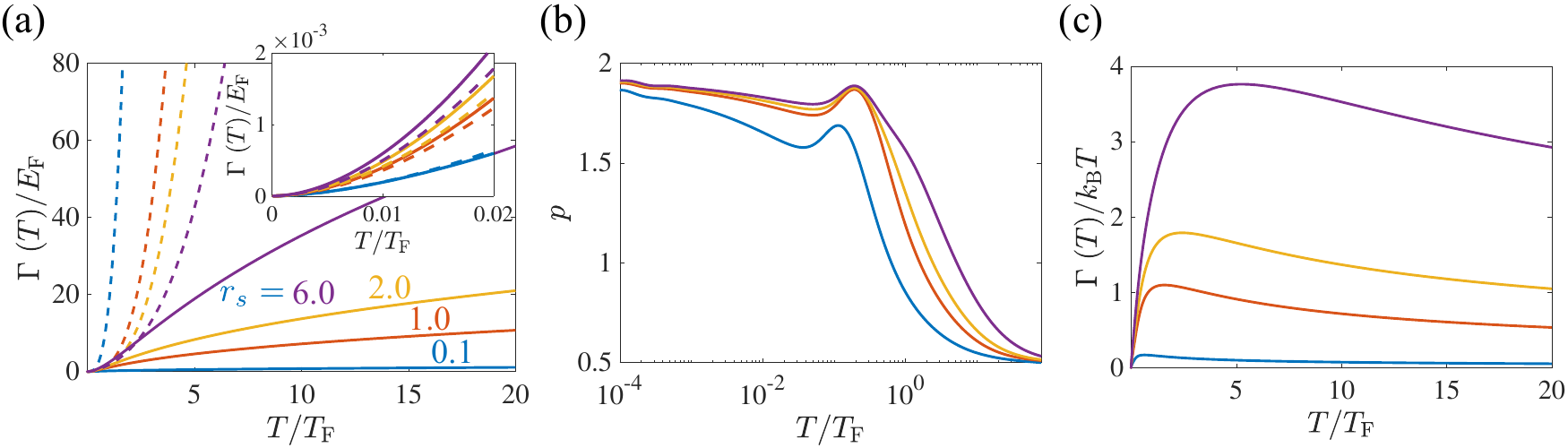}
  \caption{(a) Finite temperature scattering rates [$\Gamma(T)=\hbar/\tau_\mathrm{ee}(T)$] arising from electron-electron Coulomb interaction numerically calculated within RPA approximation along with the low-$T$ analytical asymptotic results (dashed line) for various values of $r_s=0.1$, $1.0$, $2.0$, and $6.0$. The inset shows the zoom-in of the low temperature region, showing that the numerical and asymptotic results are in good agreement. (b) The power law exponent $p$ of the numerically calculated scattering rates in (a), which saturates from $2$ to $0.5$ with increasing $T$ from $0$ to $20T_\mathrm{F}$ regardless of the value of $r_s$. (c) The Planckian parameter obtained using the results in (a). Here we use the bare electron mass with unity dielectric constant ($\kappa=1$). }  
  \label{fig:7}
\end{figure*}

In this section, we investigate the Planckian conjecture for the inelastic scattering rate arising from electron-electron Coulomb interaction. 
This is unrelated to any transport considerations in our systems, and the motivation is that inelastic scattering from electron-electron interactions may be a natural quantity for Planckian considerations \cite{hartnollPlanckianDissipationMetals2021a}.
We use the RPA approximation involving the infinite series of polarization bubble diagrams (Fig.~\ref{fig:self_energy_diagram_figure})  to evaluate the imaginary part of the self-energy, which is given by 
\begin{align}
    \mathrm{Im}\Sigma(\bm k, \omega,T)
    \!=&\!\int\!\frac{d^2 q}{(2\pi)^2} \left [n_\mathrm{B}(\hbar\omega-\xi_\mathrm{\bm k+\bm q}) + n_\mathrm{F}(-\xi_\mathrm{\bm k+\bm q}) \right ] \nonumber \\ 
&\times v_c(\bm q)\mathrm{Im}\left[\frac{1}{\varepsilon(\bm q,\xi_\mathrm{\bm k+\bm q}-\hbar\omega, T)}\right],
    \label{eq:imag_self_energy}
\end{align}
where $\xi_\mathrm{\bm k}=\varepsilon_\mathrm{\bm k}-\mu(T)$ and $n_\mathrm{F(B)}$ denotes the Fermi (Bose) distribution function. It should be noted that the Coulomb interaction is dynamically screened by the screening function $\varepsilon(\bm q, \omega, T)=1-v_c(q)\Pi(q,\omega, T)$ varying as a function of the frequency $\omega$, whereas the transport calculation involves only the static dielectric function with no dependence on $\omega$. Similar to the static case, we calculate the finite temperature dynamic polarizability $\Pi(q,\omega, T)$ using the zero temperature dynamic polarizability given by \cite{Stern1967}
\begin{align}
    \Pi(\bm q,\omega)=&-\frac{m}{\pi} + \frac{m^2}{\pi q^2}
    \left[
    \sqrt{\left(    \omega+\frac{q^2}{2m}   \right)^2-\frac{2E_\mathrm{F} q^2}{2m}}\right.\nonumber \\
    &-
    \left.\sqrt{\left(    \omega-\frac{q^2}{2m}   \right)^2-\frac{2E_\mathrm{F} q^2}{2m}}\right],
    \label{eq:iso_polar}
\end{align}
and Eq.~(\ref{eq:pol_finite_T}). 
We note that the self-energy of Eq.~(\ref{eq:imag_self_energy}), as detailed in Fig.~\ref{fig:5}, is the leading-order self-energy in the dynamically screened electron-electron Coulomb interaction, which is exact in the high-density limit.
Within the on-shell approximation, the inelastic Coulomb scattering rate at the Fermi surface at finite temperature $T$ is given by
\begin{equation}
    \hbar/\tau_\mathrm{ee}(T)=2\mathrm{Im}\Sigma(\bm k_\mathrm{F}, \xi_\mathrm{\bm k_\mathrm{F}},T).
\end{equation}
It has recently been shown that at low temperatures, $T\ll T_\mathrm{F}$, the Coulomb scattering rate asymptotically analytically behaves as \cite{liaoTwodimensionalElectronSelfenergy2020} 
\begin{align}
        \hbar/\tau_\mathrm{ee}(T)=& 
            \frac{\pi}{4}\frac{T^2}{T_\mathrm{F}}\ln{ \frac{\sqrt{2}r_s T_\mathrm{F}}{T} }\nonumber \\ 
            +& \frac{\pi}{12}\left(6 + \ln{2\pi^3} -36 \ln{A}\right)\frac{T^2}{T_\mathrm{F}}\nonumber \\ 
            -&\frac{7\zeta(3)}{\sqrt{2}\pi} \frac{T^3}{r_s T_\mathrm{F}}
            \label{eq:low_T_self_energy}
\end{align}
exhibiting the well known $\hbar/\tau(T)\sim T^2 \ln{T}$ behavior in the $T\ll T_\mathrm{F}$ limit \cite{zhengCoulombScatteringLifetime1996}. Here, $\zeta(s)$ is the Riemann-zeta function, $A=1.28243$ is the Glaisher’s constant, and $r_s$ is the dimensionless Coulomb interaction parameter characterizing the interaction strength defined as $r_s=\kappa m e^2/\hbar^2\sqrt{\pi n}$ with $n$ being the carrier density. The RPA theory is exact in the high-density or equivalently low-$r_s$ limit.

Figure~\ref{fig:7}(a) presents the numerically calculated scattering rate as a function of temperature for various values of $r_s$ along with the low temperature asymptotic curves [Eq.~(\ref{eq:low_T_self_energy})], showing good agreement between the full numerical and asymptotic results. In Fig.~\ref{fig:7}(b) we plot the power-law exponent of the scattering rate numerically calculated using $p=d\ln{\Gamma}/d\ln{T}$. Note that $p$ varies from 2 to 0.5 with increasing $T$ from $0$ to $20T_\mathrm{F}$, implying that the Planckian parameter defined as $\Gamma/k_\mathrm{B} T$ linearly increases as a function of $T$ for $T\ll T_\mathrm{F}$, and decreases as $1/\sqrt{T}$ for $T\gg T_\mathrm{F}$, resulting in local maxima in the intermediate temperature regime around $T\sim T_\mathrm{F}$, as shown in the Fig.~\ref{fig:7}(c) presenting the Planckian parameter for various values of $r_s$. It should be noted from Fig.~\ref{fig:7}(c) that the maximum of the Planckian parameter is larger for stronger interaction (i.e., larger $r_s$), but is of the order of only $2$--$4$ approximately obeying the Planckian hypothesis for the typical range of $r_s$ values of usual two-dimensional materials ($r_s\lesssim6$). 
For much larger $r_s$, i.e., very strongly interacting systems with $r_s\gg 1$, the dimensionless Planckian parameter exceeds unity by an increasingly larger factor, but the RPA theory becomes increasingly unreliable quantitatively for large $r_s$, and we are not sure whether any significance should be attached to our theory for $r_s\gg1$.
We mention that our RPA theory is precisely the leading order 1/$N$ theory in quantum field theories, where N is the number if fermion flavors, which turns out to be equivalent to the leading-order theory in $r_s$ also for an interacting Fermi liquid.

We point out that just as the asymptotically low-$T$ ($\ll T_\mathrm{F}$) analytical behavior of the 2D inelastic scattering rate goes as $T^2 \ln{T}$ (Eq.~\ref{eq:low_T_self_energy}), it is easy to show that high-$T$ ($\gg T_\mathrm{F}$) behavior goes as linear in $T^{1/2}$, simply because at high temperatures ($\gg T_\mathrm{F}$) the Fermi-Dirac statistics for the electrons becomes a Maxwell-Boltzman statistics. Thus, both our low-$T$ and high-$T$ behaviors of the numerically calculated inelastic scattering rates agree precisely with theoretical analytical expectations.

We should add that our use of the mass-shell self-energy approximation within the RPA diagrams (of Fig.~\ref{fig:5}) allows us to neglect the inclusion of any renormalization factor $Z$ in the calculation of the scattering rate, which, in principle, should be the energy-width of the quasiparticle spectral function in the context of Planckian considerations \cite{hartnollPlanckianDissipationMetals2021a}. It turns out that the calculation of the self-energy in the leading order infinite ring diagram approximation is more consistent with the mass-shell self-energy along with neglecting the renormalization factor because this provides an approximate cancellation of the higher-order diagrams in the theory \cite{riceEffectsElectronelectronInteraction1965, tingEffectiveMassFactor1975, leeLandauInteractionFunction1975}.
The full 2D interacting spectral function has recently been calculated in depth and the quantitative difference between inclusion or not of the $Z$-factor is less than a factor of 2, and, therefore, for our Planckian considerations, whether the renormalization factor is included or not is unimportant \cite{ahnFragileStableTwodimensional2021}. 

\begin{figure}[!htb]
  \centering
  \includegraphics[width=\linewidth]{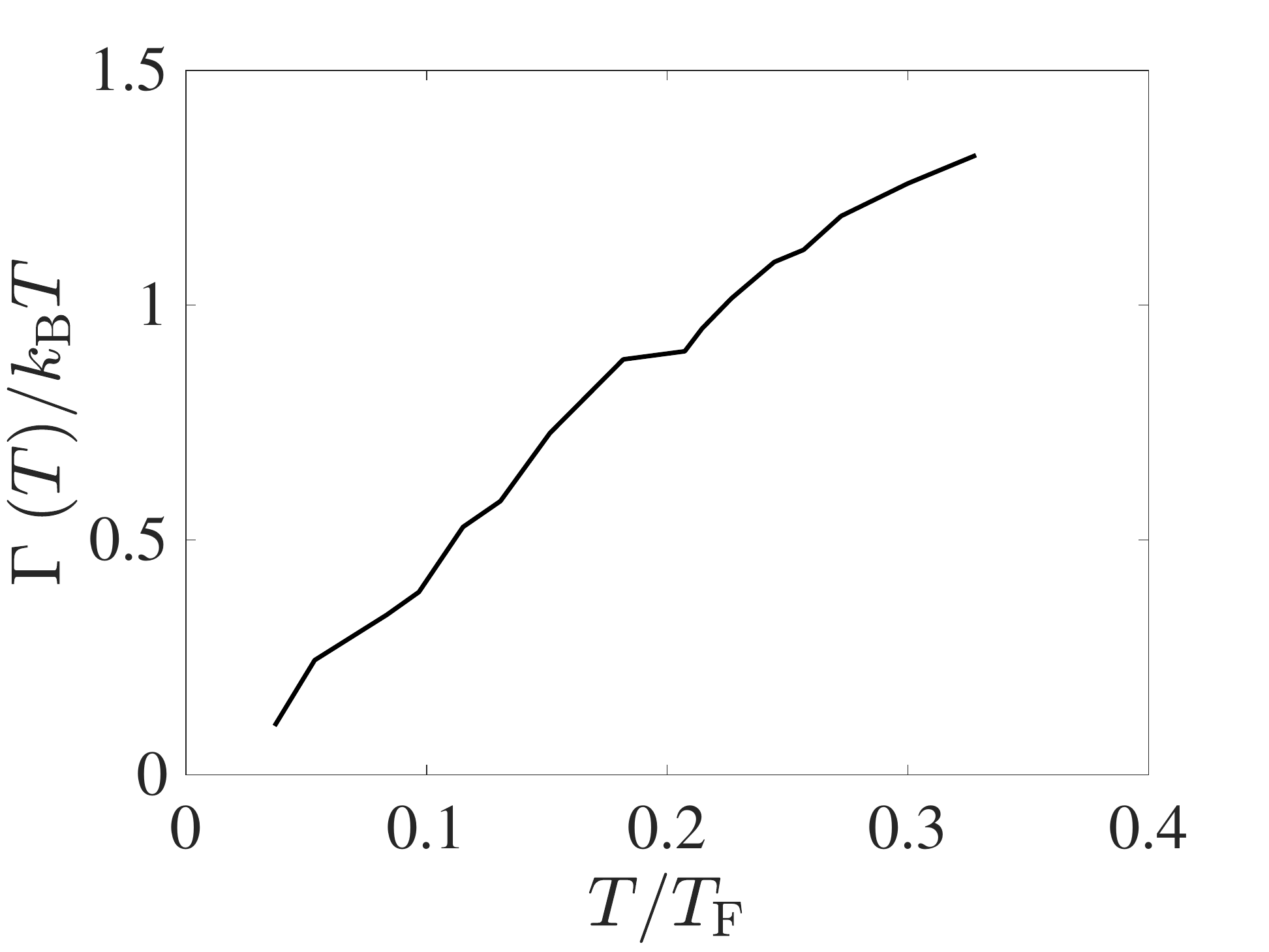}
  \caption{Experimental Planckian parameters obtained using the experimentally measured Coulomb scattering rates for GaAs present in Fig.~3 of Ref.~\cite{zhengCoulombScatteringLifetime1996, murphyLifetimeTwodimensionalElectrons1995}. Here $\Gamma(T)=\hbar/\tau(T)$, where $\tau(T)$ is the Coulomb lifetime. }  
  \label{fig:self_energy_exp}
\end{figure}

We conclude this section by presenting experimental Planckian parameters for the Coulomb scattering rate in Fig.~\ref{fig:self_energy_exp}, which we obtain using the lifetime of 2D GaAs measured through inelastic tunneling spectroscopy in Ref.~\cite{murphyLifetimeTwodimensionalElectrons1995}. It is important to note that the Planckian parameter linearly increases with increasing $T$ obeying the Planckian hypothesis with $\hbar/\tau\lesssim k_\mathrm{B} T$, in agreement with our theoretical Planckian analysis discussed in this section. These experimental inelastic tunneling scattering rates approximately agree with the RPA theory used in our work \cite{zhengCoulombScatteringLifetime1996}.

\section{ Dimensional analysis} \label{sec:5}

Our experimental analyses (Sec.~\ref{sec:2}) as well as theoretical analyses involving transport (Sec.~\ref{sec:3}) establish the approximate validity of generalized (i.e., within an order of magnitude) Planckian bounds on scattering rates in 2D semiconductors. We discuss some intuitive dimensional arguments, which are neither rigorous nor complete, but should serve as a motivation for future thinking on Planckian, which is currently an empirical finding.

As for the temperature-dependent inelastic scattering rate due to electron-electron interactions itself, as mentioned already above, energy-time uncertainty relation provides a crude `explanation' since the Fermi surface becomes `diffuse' by $k_\mathrm{B}T$ at finite temperatures, any energy uncertainty arising from inelastic scattering should be approximately bounded by $k_\mathrm{B} T$, leading to $\Gamma_\mathrm{ee} < k_\mathrm{B}T$ as found in Sec.~\ref{sec:4} above. Obviously such uncertainty relation induced inequality is only dimensionally applicable, and can at best be valid within an order of magnitude.

The above uncertainty argument, however, says nothing about transport, unless the resistive scattering itself arises from electron-electron interaction induced inelastic scattering, which is most certainly not the case for the systems of our interest in the current work (and also not for phonon-induced generic linear-in-$T$ resistivity in all normal metals).  We believe that transport measurements themselves can only extract a momentum relaxation rate associated with resistive scattering, which in general says nothing about any underlying inelastic scattering.  So, we need a different heuristic argument in order to understand the empirical validity (within an order of magnitude) of the Planckian bound for the temperature-dependent transport in doped semiconductors.

One possibility, already alluded to in Sec.~\ref{sec:3} analytical arguments, is that the approximate transport Planckian bound arises from the combination of two complementary definitions of a metal: Existence of coherent quasiparticles (Ioffe-Regel-Mott) criterion, $\Gamma < E_\mathrm{F}$, where $\Gamma= \hbar/2\tau$ with $\tau$ as the resistive momentum relaxation rate and the thermal constraint that temperature $k_\mathrm{B} T$ should not exceed the Fermi energy by too large a factor in a quantum metal.  $E_\mathrm{F}$ is given in 2D (for a circular Fermi surface) by:
\begin{equation}
    E_\mathrm{F} = \frac{\pi n \hbar^2}{m},
\end{equation}
which, combined with Eq.~(\ref{eq:IRM}) for the resistivity immediately gives:
\begin{equation}
    \rho < \frac{\hbar}{e^2}.
\end{equation}
This is of course the Ioffe-Regel-Mott criterion limiting the 2D resistivity below the resistance quantum $h/e^2$ for the existence of a metal. Now, we write the scattering rate quite generally as
\begin{equation}
\hbar/\tau = \alpha k_\mathrm{B}  T,  
\end{equation}
leading immediately to [using Eq.~(\ref{eq:finite_temperature_resistivity})]: 
$\rho = \pi  \hbar^2 \alpha  k_\mathrm{B} T /(\hbar e^2 E_\mathrm{F}) < h/e^2$, which then gives, after simplifying both sides:
\begin{equation}
    E_\mathrm{F} > \frac{\alpha  k_\mathrm{B} T} {2\pi}.
\end{equation}
If we now demand that a quantum metal is constrained to $E_\mathrm{F} < k_\mathrm{B} T$, we get:
\begin{equation}
    \alpha < 2\pi,     
\end{equation}
which is essentially our generalized Planck bound for transport. 
While we do not claim this line of reasoning to be anything more than a heuristic dimensional argument, it is consistent with the finite temperature transport scattering rate to be constrained from above by a $\mathcal{O}(1- 10)$ based simply on the requirement that the electron system preserves a well-defined Fermi surface (which is the fundamental definition of a metal) both against momentum decoherence and thermal broadening. This dimensional argument hints that the empirical validity of the Planckian bound may simply be a manifestation of the dual facts that the natural scales for the quasiparticle momentum and energy in a Fermi system are Fermi momentum and Fermi energy, and this leads to a natural scale for the transport scattering rate being of the order of $k_\mathrm{B} T$.  It would be meaningful to try to make this dimensional argument more rigorous although that may turn out to be a challenge in strongly-coupled  disordered interacting systems with no natural small for any perturbative expansion to work.

\section{Conclusions} \label{sec:6}

We find that the strong observed temperature dependence of the low-temperature ($<10K$) resistivity in high-quality 2D doped semiconductors to follow, both experimentally (Sec.~\ref{sec:2}) and theoretically (Sec.~\ref{sec:3}), sub-Planckian ($\hbar/\tau < k_\mathrm{B}T$), Planckian ($\hbar/\tau \sim k_\mathrm{B}T$), and super-Planckian ($\hbar/\tau>k_\mathrm{B} T$) behaviors as a function of carrier density, with the super-Planckian behavior manifesting at  lower carrier densities where the temperature dependence of the resistivity is the strongest and the resistivity itself is the largest (but still below the Ioffe-Regel-Mott limit to ensure metallicity). Two noteworthy features of our results are that (1) the violation of the Planckian bound in the super-Planckian regime is always rather modest, within a factor of $2$--$6$ (i.e., of the order of 10 or less), and (2) the super-Planckian behavior manifests not only at lower carrier densities, but also at a rather high effective dimensionless temperature ($T/T_\mathrm{F} \sim 1/3$ or so, a regime totally inaccessible in normal metals)—in fact, for all densities the dimensionless Planckian parameter ($\hbar/\tau$)/($k_\mathrm{B}T$) is at its maximum at a finite temperature $\sim T/T_\mathrm{F} \sim 1/3$, just before the quantum-classical crossover in the resistivity leading to a decreasing resistivity ($\sim 1/T$) with increasing temperature. The temperature dependence here arises not from electron-phonon interactions (which would be operational in these systems for $T>10K$--$20K$ typically) or from umklapp or Baber electron-electron scattering, but from a combination of disorder and indirect electron-electron interactions through the screening of the disorder (or equivalently from the temperature dependent Friedel oscillations). We emphasize that the temperature dependence disappears if the electron-electron interaction is set to zero, thus using unscreened disorder in the theory.

Our empirical findings based on a detailed quantitative analysis (Sec.~\ref{sec:2}) of the temperature and density dependent experimental 2D transport data in different materials are supported by our detailed transport theory (Sec.~\ref{sec:3}) based on resistive scattering by finite-temperature momentum-dependent screened Coulomb disorder (or equivalently by the Friedel oscillations associated with the screened charged impurity potential). The temperature-dependence in fact goes away if the theory uses long-wavelength Thomas-Fermi screening without the 2$k_\mathrm{F}$-kink in the $T=0$ 2D polarizability.

While the scattering rate $1/\tau$ for transport is a momentum relaxation rate, we also consider (Sec.~\ref{sec:4}) theoretically the temperature- and density-dependent inelastic scattering rate $1/\tau_\mathrm{ee}$ arising from the many-body electron-electron Coulomb interaction, which is momentum-conserving in our theory since a doped semiconductor does not allow umklapp scattering. The inelastic electron-electron interaction scattering rate is simply given by the imaginary part of the 2D self-energy, and is conceptually qualitatively different from the $1/\tau$ transport scattering rate defining the resistivity in the transport theory (Sec.~\ref{sec:3}). We find that $1/\tau_\mathrm{ee}$ also obeys the Planckian bound of being within a factor of $10k_\mathrm{B} T$ in general, with the peak value of the dimensionless Planckian parameter, ($\hbar/\tau_\mathrm{ee}$)/($k_\mathrm{B} T$), also occurring around $T\sim T_\mathrm{F}$ as in the transport problem. We find it intriguing that both our (energy-conserving) screened disorder induced transport scattering rate and our (momentum-conserving) interaction induced inelastic scattering rate approximately obey the Planckian bound, with the super-Planckian behavior being only a modest factor of $2$--$6$, happening only at lower densities and higher temperatures, i.e., around $T\sim T_\mathrm{F}$ (a lower density implies a lower $T_\mathrm{F}$). We emphasize that this approximate Planckian behavior we discover in 2D doped semiconductor properties applies to both theory (Secs.~\ref{sec:3} and \ref{sec:4}) and experiment (Secs.~\ref{sec:2} and \ref{sec:4}). Our Planckian analysis of the experimental data establishes an empirical applicability of the Planckian hypothesis to 2D semiconductor transport with a modest super-Planckian behavior at low densities around $T\sim T_\mathrm{F}$, and our theory, for both the momentum relaxation rate and the inelastic scattering rate, indicates that the Planckian hypothesis applies (within an order of magnitude) theoretically over essentially all densities and temperatures in the metallic regime for 2D semiconductors. Why? Why is the Planckian conjecture apparently approximately valid in our systems (involving different 2D semiconductor materials) over all metallic ranges of density/temperature?

The answer to the question of why the Planckian bound applies, even approximately, to any experimental (or even physical) properties is unclear at this point in spite of many experimental reports of its empirical validity in transport properties of many materials, starting with the important observation by Bruin, et al \cite{bruinSimilarityScatteringRates2013}. The reverse is, in fact, true\texttt{--}a compelling recent theoretical analysis by Lucas \cite{lucasOperatorSizeFinite2019}\cmmnt{[https://doi.org/10.1103/PhysRevLett.122.216601]} shows that no such bound should exist for the temperature-dependent resistivity.
In particular, temperature defines thermodynamic equilibrium whereas transport is necessarily a dissipative kinetic property, where temperature enters as a parameter defining the distribution functions, so why there should be a fundamental transport bound defined by $k_\mathrm{B} T$ is unclear.
It is also an obvious fact that the bound does not apply to the residual disorder-induced resistivity at $T=0$ that all metals (ignoring any superconductivity) must have.  It is therefore suggested \cite{hartnollPlanckianDissipationMetals2021a}\cmmnt{[arXiv:2107.07802]} that any Planckian bound must not include any elastic scattering since at $T=0$ only elastic scattering (by disorder) can contribute to transport.  This is, however, in sharp contrast with the most established empirical Planckian behavior observed routinely in normal metallic resistivity ($> 40$K), which was already emphasized in Ref.~\cite{bruinSimilarityScatteringRates2013}\cmmnt{ [1]} and has been well-known since the 1950s \cite{ziman1972principles}\cmmnt{ [Ziman, book]}, where a linear-in-$T$ resistivity ranging from sub-Planckian (e.g., Al) to super-Planckian (e.g., Pb) manifests itself arising from the quasi-elastic electron-phonon scattering in the equipartition regime above the Bloch-Gruneisen temperature \cite{hwangLinearinResistivityDilute2019}\cmmnt{ [cite Hwang-SDS Linear-in-T PRB 2019]}.  The Planckian behavior of phonon-induced metallic resistivity follows from the simple fact that the corresponding scattering time, $\tau_\mathrm{ep}$, follows the simple formula
\begin{equation}
    \hbar/\tau_\mathrm{ep} = 2\pi \lambda k_\mathrm{B} T, 
\end{equation}
in the phonon equipartition regime, with $\lambda$ being the dimensionless electron-phonon Eliashberg-McMillan coupling constant.  It just so happens that $\lambda$ in common metals lie typically between $0.1$ and $1.5$. But, even for this well-established phonon scattering induced linear-in-$T$ metallic resistivity, the Planckian bound seems to apply approximately (within a factor of 10) for all metals since the metallic electron-phonon coupling constant seems empirically to never exceed 1.5, thus ensuring that the Planckian bound is obeyed always within a factor of $10$ or so in all metals. It has been emphasized that the phonon-induced high-temperature electronic resistivity could, in principle, be anything as long as the coupling is large \cite{hwangLinearinResistivityDilute2019}\cmmnt{ [Hwang-SDS 2019]} with the transport theory imposing no bound at all, but empirically this does not seem to happen with the electron-phonon coupling never being much larger than unity ever.  Why such a Planckian bound applies for the electron-phonon coupling remains unclear although the fact that this happens is empirically well-established. It has recently been speculated that the observed empirical limit on the electron-phonon coupling (i.e., $\lambda < 1.5$) may be related to a materials stability bound \cite{murthyStabilityBoundLinear2021}\cmmnt{ [ arXiv:2112.06966]}. This speculated lattice stability bound, however, has no relevance to our findings in the current work where phonons do not play any role.

The strongly temperature-dependent resistive scattering in doped 2D semiconductors considered in our work arises from a temperature-dependent effective disorder although the bare disorder, arising from random quenched charged impurities, is temperature independent, with the temperature dependence in the effective disorder arising from the anomalously strongly temperature-dependent finite-momentum 2D polarizabilty function controlling the screening of the bare disorder.  The scattering in our case is an energy-conserving elastic scattering, as appropriate for quenched disorder, causing a momentum relaxation in spite of the strong temperature dependence.  
We emphasize that the temperature dependence disappears in our theory if we suppress the electron-electron interaction or the electronic polarizability so that the disorder is unscreened, and thus, the temperature dependence arises entirely from the indirect effects of interaction.
The significant point is that the temperature dependence of resistivity here, in spite of arising from elastic scattering, is strong and asymptotically linear down to arbitrarily low temperatures, with nothing strange or non-Fermi-liquid about it except in the sense of a trivial non-Fermi liquid as defined in \cite{buterakosPresenceAbsenceTwodimensional2021a}\cmmnt{ [Buterakos-Das Sarma PRB 2020]}. This should serve as a cautionary note to numerous experimental claims in the literature that a linear-in-$T$ resistivity extending to low temperatures necessarily implies a non-Fermi liquid strange metal. The high-quality low-density doped 2D semiconductors are a manifest counterexample to such strange metal assertions.

The Planckian properties of our systems have some similarities and some differences with phonon-induced Planckian behavior of normal metals: (1) Both resistive scattering are energy-conserving and quasi-elastic; (2) both manifest strong temperature dependence; (3) both our systems and metals manifest sub-Planckian, Planckian, and super-Planckian behaviors, depending on the metal (i.e., the $\lambda$-dependence), and depending on the materials system and carrier density of the 2D semiconductor (i.e., the $q_\mathrm{TF}/2k_\mathrm{F}$-dependence); (4) the $T$-dependence in our case is induced by 2D screened disorder in contrast to normal metals where the $T$-dependence is induced by phonon scattering in the equipartition regime; (5) our $T$-dependence is a low-$T$ phenomenon set by the Fermi temperature ($T<T_\mathrm{F}$) whereas the $T$-dependence in normal metals is a high-$T$ ($>T_\mathrm{BG}$) phenomenon set by the Bloch-Gruneisen ($T_\mathrm{BG}$) or Debye temperature; (6) the phonon-induced $T$-dependent resistivity is essentially linear in the equipartition regime $T>T_\mathrm{BG}$, whereas our 2D resistivity is linear only for low temperatures, $T\ll T_\mathrm{F}$.

We mention that the Planckian bound, to the extent it applies, should not be constrained to metals manifesting just an approximate $T$-linear resistivity, since such a linear constraint is meaningless in solid state physics as no resistivity can really be precisely linear (the power law may be very close to unity over a limited range of $T$, but never precisely unity). 
In fact, careful recent resistivity measurements in graphene layers show that the strange Planckian behavior persists for temperature power laws different from unity \cite{jaouiQuantumCriticalBehavior2022}. 
Although the Planckian bound is often discussed in the context of `strange metals’ manifesting a large linear-in-$T$ resistivity over a range of temperature, the two concepts are distinct and have nothing to do with each other (except perhaps in a negative sense in that the most established generic example of a Planckian behavior is found in the linear-in-$T$ resistivity of normal metals, a most non-strange situation, arising from phonon scattering in the equipartition regime \cite{wuPhononinducedGiantLinearin2019}\cmmnt{ [ cite Wu, Hwang, SDs ‘ordinary strangeness PRB and Hwang-SDS linear-in-T PRB])}). Thus, we define the Planckian bound as a constraint on $\tau(T)$, where $\hbar/\tau(T) \lesssim \alpha k_\mathrm{B} T$, where $\alpha$ is a dimensionless number of order $1$--$10$, independent of whether the scattering time $\tau$ is strongly or weakly dependent on $T$. 
In our case (Secs.~\ref{sec:2} and \ref{sec:3}), the resistivity is indeed linear in the asymptotic low-$T$ regime $T\ll T_\mathrm{F}$, but in reality the low-density 2D systems have rather low values of $T_\mathrm{F}$, and therefore the temperature-dependent resistivity often departs from linearity with increasing $T$ (eventually decreasing with increasing $T$ beyond the quantum-classical crossover regime). Our theory for the inelastic scattering has the explicit analytical form that $1/\tau_\mathrm{ee}$ goes as $T^2 \ln T$ ($T^{1/2}$) for $T\ll(\gg)T_\mathrm{F}$, but the Planckian hypothesis seems to apply at all temperatures approximately.

In the recent Planckian literature, a great deal of significance is placed on the coefficient $\alpha$ being unity (i.e., defining a mysterious and mystical entity, the so-called Planckian metal). We do not agree with this assertion of $\alpha=1$ being of deep significance as our results show that $\alpha$ could vary from being $\ll 1$ to being $\sim 1$--$10$ in the same sample depending on the carrier density (which cannot be varied in situ in a single sample in strongly correlated materials, each sample coming with its own fixed, and essentially unknown, carrier density). If we fine tune and choose some narrow post-selected density and temperature range, we get an effective $\alpha \sim 1$, but this is simply confirmation bias of varying parameters until we find what is being looked for. In addition, $\alpha=1$ as a strict requirement for a Planckian metal defines a set of measure zero since there is no claim that $\alpha$ is somehow an invariant. So any measured $\alpha$ would necessarily depart from unity, particularly since obtaining $\alpha$ from transport measurements necessitates a precise knowledge of $n/m$, which is simply unavailable in strongly correlated materials. Although we do not believe that much significance can be attached to the recent claims about $\alpha$ precisely being unity as being special, we are surprised that our results for 2D semiconductors exhibit that, although $\alpha$ can be much less than unity (``sub-Planckian''), it is never much larger than unity with the super-Planckian behavior being constrained by $\alpha <10$ always. Why?

In this context, we mention that in direct conflict with recent fine-tuned claims of hole-doped cuprates having $\alpha\sim 1$, and thus being strange Planckian metals, older direct optical measurements of $\alpha$ give results consistent with our conclusion that generically $\alpha$ could take values in the range of $1-3$ in the super-Planckian regime with there being nothing special about $\alpha \sim 1$ except in a fine-tuned sense. These optical measurements give $\alpha= 1.57$ (LSCO), $2.5$ (YBCO), $1.97$ (BSCCO) \cite{gaoInfraredPropertiesEpitaxial1993, gaoQuasiparticleDampingCoherence1996, romeroQuasiparticleDampingBi1992}.
In addition, recent measurements in electron-doped cuprates have reported \cite{poniatowskiCounterexampleConjecturedPlanckian2021} 
$\alpha \sim 1$--$3$ in the super-Planckian regime, again showing, consistent with our finding, that any empirical bound on $\alpha$ is $\lesssim 10$, and not 1. This is also consistent with normal metals where the largest value (for Pb) of $\alpha$ is $\alpha \sim 9$, although many metals exhibit sub-Planckian behavior with $\alpha< 1$.
We believe that our work indicates that the theoretical focus should shift to why $\alpha$ is not arbitrarily large and remains bounded approximately by $10$ rather than focusing on the misleading fine-tuned claims of $\alpha \sim 1$ being a generic strange metal of particular interest \cite{grissonnancheLinearinTemperatureResistivity2021, legrosUniversalTlinearResistivity2019,caoStrangeMetalMagicAngle2020, yuanScalingStrangemetalScattering2022a}.

The empirical validity of the generalized Planckian conjecture (with the super-Planckian violation of the bound being always less than 10) in the transport properties of our systems (Secs.~\ref{sec:2} and \ref{sec:3}) as well as in other systems \cite{hartnollPlanckianDissipationMetals2021a}\cmmnt{ [Hartnoll review arXiv]} remain a mystery, and the possibility that it is a coincidence cannot be ruled out (We have ruled out the bound being unity, i.e. $\alpha=1$, as a fine-tuned confirmation bias since $\alpha > 1$, but $<10$, appears to be the generic super-Planckian behavior in many systems, including the ones studied in the current work\texttt{--} of course, the sub-Planckian $\alpha < 1$ situation is the most generic situation). One theoretical bound, which has attracted considerable attention, is associated with quantum chaos and relates a Lyapunov exponent, $1/\tau_\mathrm{L}$, for the rate of growth of chaos in thermal quantum systems with a large number of degrees of freedom \cite{maldacenaBoundChaos2016}\cmmnt{ [arXiv:1503.01409]}:
\begin{equation}
    \hbar/\tau_\mathrm{L} \lesssim  k_\mathrm{B} T.
\end{equation}
This bound is connected with the fact that chaos can be diagnosed using an out-of-time-order correlation (OTOC) function closely related to the commutator of operators separated in time \cite{larkinQuasiclassicalMethodTheory1969}.  But OTOC and the corresponding Lyapunov exponent $1/\tau_\mathrm{L}$ do not have any established (or even speculated) relationship with the transport scattering rate, and therefore, the implication of this conjectured OTOC bound to Planckian transport bound considerations is unclear.\cite{xuButterflyEffectInteracting2019}\cmmnt{ [cite Xu,…Swingle, SDS butterfly paper]}.  Thus, the empirical validity of the (approximate or generalized) Planckian transport bound is not explicable by quantum chaos considerations. There are speculations, none convincing at all, that the Planckian conjecture may be related to the holographic viscosity bound in quantum field theories, but the applicability of such holographic duality to concrete experimental condensed matter problems is purely speculative and unconvincing \cite{kovtunViscosityStronglyInteracting2005}. 

One can construct artificial theoretical models which lead to linear-in-$T$ strange metal Planckian behaviors simply by embedding this physics intrinsically into the model without any microscopic rationale. The most-well-known such model is the so-called `marginal' Fermi liquid model \cite{varmaPhenomenologyNormalState1989},
where one just assumes ad hoc that the imaginary part of the electron self-energy goes as linear in $T$, and all scattering is umklapp, breaking momentum conservation, so that the scattering rate goes precisely as $ k_\mathrm{B} T$, and the system is marginally not a Fermi liquid. This is, however, assuming the result we want, and in spite of more than 30 years of efforts since the original introduction of this marginal Fermi liquid model in the context of the hole-doped cuprates, no microscopic theoretical justification exists for how such a singular self-energy, in contrast to the well-known $T^2\ln{T}$ behavior, could arise in 2D and why the corresponding resistivity should be given precisely by this imaginary part of the single-particle self-energy ignoring all momentum dependence in the problem.  Without any microscopic justification, such a marginal Fermi liquid model is simply an assertion, and not a theory, and cannot be taken seriously in the context of experimental findings.
There was important theoretical work establishing non-Fermi-liquid behaviors of 2D fermions coupled to charged black holes \cite{leeNonFermiLiquidCharged2009}
using AdS/CFT correspondence \cite{maldacenaLargeNLimitSuperconformal1999} 
as well as 2D fermions coupled with U(1) gauge fields \cite{leeLowenergyEffectiveTheory2009}. 
While these works are important proofs of principle for the theoretical existence of 2D non-Fermi liquid behavior, their connections to any physical 2D materials systems in condensed matter physics have remained completely elusive in spite of many efforts.
There is impressive recent theoretical work \cite{patelUniversalLinearResistivity2022}
establishing that special classes of theoretical  quantum critical 2D metals with completely spatially random fluctuations in the fermion-scalar Yukawa couplings may lead to linear-in-$T$ resistivity by construction, but these results do not manifest any generic Planckian behavior with the resulting $\alpha\ll 1$ in general (without fine-tuning). Of course, linear-in-$T$ resistivity with $\alpha<1$ arises generically in metals with weak electron-phonon coupling and in our screened disorder coupling theory of Sec.~\ref{sec:2} for high carrier density and low temperature. Finally, the minimal Hubbard model at very high temperatures (much larger than the band-width) also leads to a linear-in-$T$ resistivity if one assumes complete momentum-independent umklapp scattering with local interactions, where each electron-electron scattering event contributes to the resistivity. But this is a highly fine-tuned trivial result appearing simply from the leading-order high-$T$ expansion of the thermal averaging, and cannot have any practical significance to real electronic materials. Thus, no generic theoretical arguments exist for a Planckian dissipative resistivity in electronic materials.

It is speculated \cite{hartnollPlanckianDissipationMetals2021a}\cmmnt{ [Hartnoll review arxiv]} that the Planckian bound may apply to inelastic scattering  rates arising from electron-electron interactions. This distinction between elastic and inelastic scattering has little significance for transport properties because transport only probes resistive scattering, which is associated with momentum relaxation, without distinguishing between elastic and inelastic scattering. In fact, as is well-known, electron-electron interaction by itself cannot relax momentum in a translationally invariant system and does not contribute to resistive scattering unless an explicit momentum conservation breaking mechanism (e.g., umpklapp scattering, Baber scattering) is invoked \cite{poniatowskiCounterexampleConjecturedPlanckian2021}.\cmmnt{[cite the recent Greene, SDS and other exptl paper]} Electron-electron interaction, however, leads to a real quasiparticle damping and finite lifetime, $\tau_\mathrm{ee}$, through the imaginary part of the self-energy, as described in Sec.~\ref{sec:4}, which may be studied experimentally by measuring the quasiparticle spectral function using inelastic tunneling spectroscopy or ARPES. We emphasize that $\tau_\mathrm{ee}$ is generically different from the transport scattering time $\tau$ obtained from the resistivity, and in general, there is no simple relationship connecting the two.

Focusing entirely on interaction-induced inelastic scattering rate (Sec.~\ref{sec:4}) from the Planckian perspective, emphasizing again that this is unrelated to the transport properties discussed in Secs.~\ref{sec:2} and \ref{sec:3}, we show that the temperature and density dependent inelastic scattering rate arising from electron-electron interaction also obeys the generalized Planckian bound approximately (i.e., within a factor of 10) for all temperature and density with a modest super-Planckian behavior manifesting around $T\sim T_\mathrm{F}$, where the scattering rate $1/\tau_\mathrm{ee}$ crosses over from the low-temperature (for $T\ll T_\mathrm{F}$)  $T^2$  behavior to the high-temperature ($T\gg T_\mathrm{F}$) $T^{1/2}$ behavior.  Thus, we find that for doped 2D semiconductor systems both the temperature dependent elastic scattering rate (arising from screened disorder) and the temperature dependent inelastic scattering rate (from electron-electron interaction) empirically obey the generalized Planckian bound approximately (i.e., is bounded, within an order of magnitude, by $k_\mathrm{B} T$)

An intuitively-appealing qualitative dimensional argument for why the inelastic scattering rate may have an approximate Planckian bound is the following. At finite temperatures, the Fermi distribution develops a thermal broadening of $k_\mathrm{B} T$ around the Fermi energy, leading to an energy uncertainty of $\mathcal{O}(k_\mathrm{B} T)$.  Since an inelastic scattering rate of $1/\tau_\mathrm{ee}$ leads, by the uncertainty relation, to an energy uncertainty $\sim \hbar/\tau_\mathrm{ee}$, it is possible that $\hbar/\tau_\mathrm{ee}$ should not exceed $k_\mathrm{B} T$ by a large number.  This uncertainty relation based qualitative argument, which is by no means rigorous, does indicate that the inelastic scattering rate $\hbar/\tau_\mathrm{ee}$ should be of $\mathcal{O}(k_\mathrm{B} T)$ or less, as we find in our detailed calculations presented in Sec.~\ref{sec:4}.

The qualitative energy uncertainty argument, however, provides no rationale whatsoever for why the transport scattering rate extracted from the resistivity, which for our systems is a temperature-dependent elastic scattering by screened disorder, should have anything at all to do with the Planckian bound  of $k_\mathrm{B} T$ as established empirically for the experimental data in Sec.~\ref{sec:2}. The only clue for a Planckian bound can be discerned from our analytical theory in Sec.~\ref{sec:3} where we explicitly used the Ioffe-Regel-Mott criterion to bound the $T=0$ 2D resistivity by a maximum possible metallic resistivity of $h/e^2$, which then leads to the following constraint for our temperature dependent transport scattering rate $1/\tau$ (with the $T=0$ contribution subtracted out):
\begin{equation}
    \hbar/\tau < 4k_\mathrm{B} T
\end{equation}
Our direct numerical results for the calculated 2D metallic resistivity are consistent with this analytical effective Planckian bound.  Thus, at least for our problem, the Planckian transport properties appear to have an underlying deep (if somewhat indirect) connection to the Ioffe-Regel-Mott criterion for the existence of a $T=0$ metal.  Since a metal with coherent quasiparticles cannot have an arbitrarily large resistivity, by definition, the temperature-dependent scattering rate cannot be arbitrarily large either, and the natural bound at finite temperature on a purely dimensional ground can only be $k_\mathrm{B} T$.  Thus, the Planckian transport bound may ultimately be arising from the fact that the mean free path in a metal at finite temperatures cannot be arbitrarily short. In our analytical theory this connection between the Planckian bound on the finite-temperature scattering rate and the Ioffe-Mott-Reel criterion is explicit, but whether the same happens for all Planckian systems (many of which are also effectively 2D) is unknown, and should be investigated in future woks.
For our problem, we establish theoretically (Sec.~\ref{sec:3}) that the Planckian conjecture holds approximately, $\hbar/\tau < 4k_\mathrm{B} T$, because of the requisite consistency with the Ioffe-Regel-Mott criterion.  This is true at low temperatures, $T\ll T_\mathrm{F}$, where $\rho (T)$ is linear in $T$, but the approximate Planckian bound continues holding all the way to $T \sim T_\mathrm{F}$ because the next-to-the-leading-order correction arising from the $(T/T_\mathrm{F})^{3/2}$ term remains small.  Whether such a direct connection to the Ioffe-Regel-Mott metallicity condition applies to other strange Planckian metals are not unknown at this point.

We provide two additional circumstantial arguments somewhat in line with the our discussion above where the Planckian bound may be connected to the very definition of a metal: (1) First, the suggestion \cite{murthyStabilityBoundLinear2021} that the electron-phonon coupling may be bounded from above (which is the reason why phonon-induced resistivity of metals obeys the Planckian conjecture) is because a much larger electron-phonon coupling would lead to a lattice instability causing a metal-to-insulator transition (which is consistent with our finding that the Planckian bound may ultimately be arising from the definition of metallicity itself); (2) second, the fact that all strongly-correlated 2D systems saturating the Planckian bound (or manifesting super-Planckian behavior) typically always have very large resistivity (this again, reinforcing that the bound may arise simply to preserve metallicity).  Much more work is needed to convert these qualitative ideas into a general theory which applies to all systems, but for our systems, we establish these ideas in Sec.~\ref{sec:3} through a detailed theory.

\begin{figure}[!htb]
  \centering
  \includegraphics[width=\linewidth]{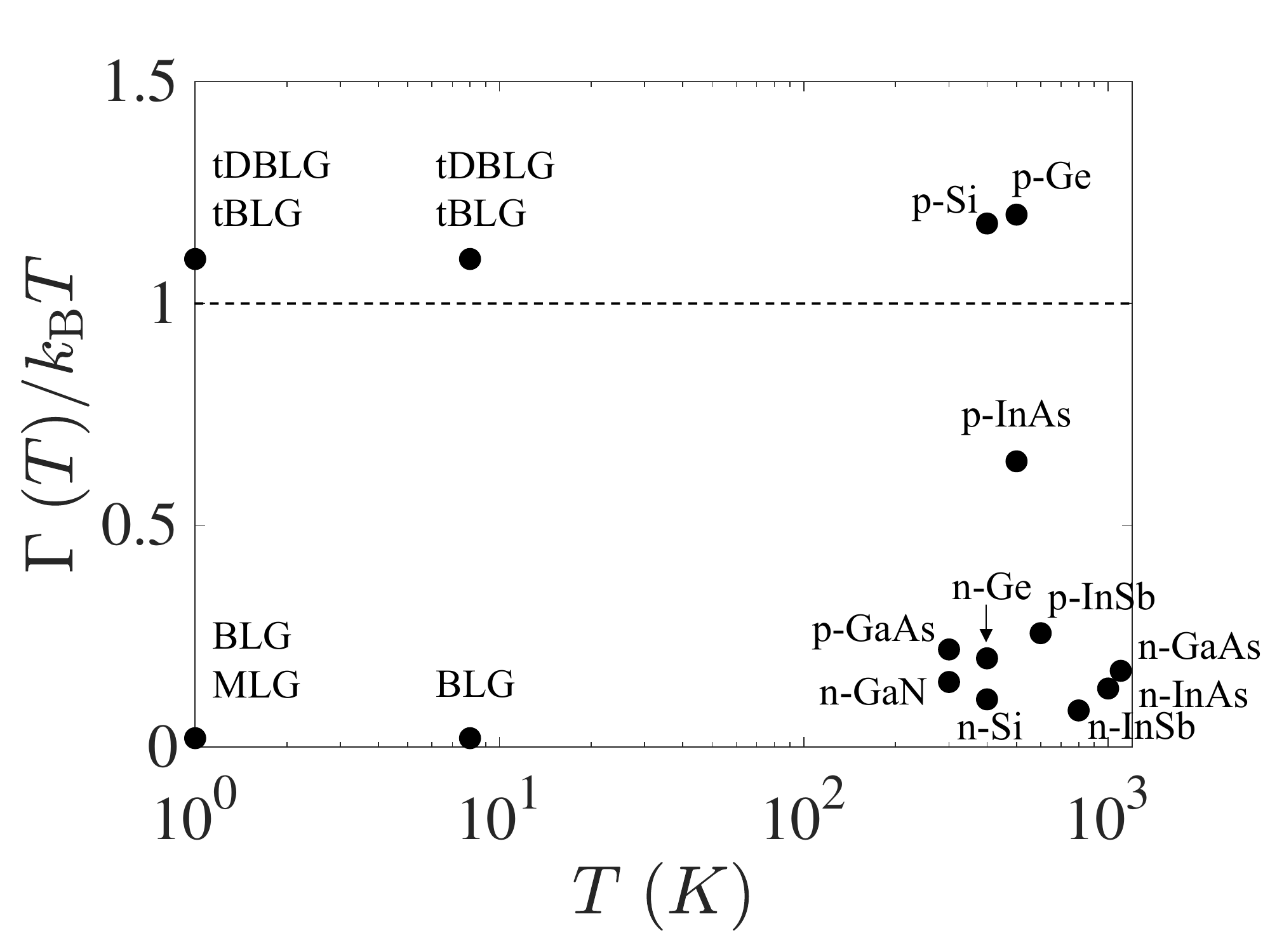}
  \caption{Planckian parameters from the experimental literature plotted against temperature for many materials, showing that the Planckian conjecture is always approximately valid.}  
  \label{fig:table_figure}
\end{figure}

We conclude by mentioning that we have also investigated the experimentally reported finite temperature transport properties of 2D graphene related systems \cite{caoStrangeMetalMagicAngle2020,jaouiQuantumCriticalBehavior2022, wuPhononinducedGiantLinearin2019, dassarmaStrangeMetallicityMoir2022,hwangAcousticPhononScattering2008, efetovControllingElectronPhononInteractions2010,minChiralitydependentPhononlimitedResistivity2011,polshynLargeLinearintemperatureResistivity2019, chuPhononsQuantumCriticality2021} and 3D doped semiconductors \cite{semicondDatabase} (at high temperatures, so that elastic scattering effects from the impurities are not particularly quantitatively important), finding to our surprise that the Planckian hypothesis is always approximately empirically valid. This is true for 3D semiconductors even at very high temperatures ($\sim1000$K) where the resistive scattering arises entirely from inelastic optical phonon scattering, and the resistivity is essentially exponential in temperature. We show these results in Fig.~\ref{fig:table_figure} where the dimensionless Planckian parameter is plotted against temperature for many materials with the Planckian conjecture being always approximately valid. Why the Planckian conjecture seems to apply for quasi-elastic acoustic phonon scattering \cite{bruinSimilarityScatteringRates2013}\cmmnt{[Bruin Ref. 1]}, inelastic optical phonon scattering, temperature-dependent screened disorder scattering (Secs.~\ref{sec:2} and \ref{sec:3}), inelastic electron-electron scattering (Sec.~\ref{sec:4}), and perhaps in many other situations still remains a mystery except perhaps for its qualitative connections to the energy uncertainty relation and the stability of the metallic phase itself.

In conclusion, we find (Secs.~\ref{sec:2} and ~\ref{sec:3}), rather unexpectedly, that the temperature dependent part of the resistivity in 2D doped semiconductors, arising from resistive scattering by temperature-dependent screened Coulomb disorder, approximately obeys the generalized Planckian hypothesis (in the sense $\alpha<10$). In addition, we find (Fig.~\ref{fig:table_figure}) the very surprising result that doped semiconductors approximately obey the Planckian hypothesis at very high temperatures where the resistive scattering arises entirely from optical phonon scattering. We also establish (Sec.~\ref{sec:4}) that the inelastic scattering induced by electron-electron Coulomb interactions obeys the generalized Planckian hypothesis at all temperatures and densities. Our work considerably expands the scope of the Planckian conjecture by establishing its surprising empirical validity to doped semiconductors, which are not considered to be either `strongly correlated' or `quantum critical'\texttt{--} in fact, doped semiconductors are basically interacting electron liquids in a jellium background with the lattice or narrow band Mott-Hubbard physics playing no role in its transport. The fact that a generalized Planckian hypothesis may apply even to a temperature-dependent resistivity arising from disorder scattering (with the temperature dependence itself happening because of anomalous screening) is particularly perplexing since there can be no Planckian constraint at $T=0$ for the residual resistivity arising from the same disorder scattering. Our finding that the generalized Planckian hypothesis does apply to the inelastic electron-electron scattering seems more plausible, but we emphasize that there is no theory predicting it and our results simply agree empirically with the Planckian bound after the fact.  None of the various theoretical attempts \cite{hartnollTheoryUniversalIncoherent2015, lucasResistivityBoundHydrodynamic2017,  hartmanUpperBoundDiffusivity2017, hanLocalityBoundDissipative2018, nussinovExactUniversalChaos2021, hanQuantumScramblingState2019,hartnollPlanckianDissipationMetals2021a,  zaanenPlanckianDissipationMinimal2019, zaanenPlanckianDissipationMinimal2019, chenManybodyQuantumDynamics2020, yinBoundQuantumScrambling2020, chenFiniteSpeedQuantum2019,lucasOperatorSizeFinite2019,xuButterflyEffectInteracting2019,maldacenaBoundChaos2016}\cmmnt{  [ e.g.,  arXiv:1405.3651 ,  arXiv:1704.07384 ,    arXiv:1706.00019 ,  arXiv:1806.01859 ,    arXiv:1812.07598 ,   arXiv:2107.07802 ,  https://doi.org/10.21468/SciPostPhys.6.5.061,   arXiv:2007.10352  ,  arXiv:2005.07558 ; https://arxiv.org/abs/1907.07637;   Phys. Rev. Lett. 122, 216601 (2019) ;   arXiv:1902.07199 ;    arXiv:1503.01409 ]} to connect the Planckian constraint to holography, hydrodynamics, quantum criticality, quantum chaos, scrambling, and other generic concepts can actually explain the unreasonable empirical validity of the generalized Planckian bound ranging from high-temperature resistivity in normal metals (i.e., acoustic phonon scattering) and semiconductors (i.e., optical phonon scattering) all the way to strongly correlated materials (i.e., some `unknown' mechanism) through the low-temperature transport in 2D doped semiconductors (i.e., screened disorder scattering). The Planckian hypothesis has a remarkable qualitative correspondence with the Ioffe-Regel-Mott criterion defining metallicity to be constrained by $\hbar/\tau < E_\mathrm{F}=k_\mathrm{B} T_\mathrm{F}$, which loosely describes the crossover from coherent metallic transport to disorder-induced strong localization (at $T=0$) or to incoherent transport (at finite $T$), but making this qualitative connection formally precise is a challenge.  This is particularly so because the Ioffe-Regel-Mott criterion is not really a sharply defined transition, it is an intuitively appealing crossover criterion defining a metal as a system with coherent quasiparticles carrying current with momentum as a reasonable quantum number except for resistive scattering events changing momenta.  It is therefore appealing that the Planckian hypothesis is in some sense the energy-time version of the Ioffe-Regel-Mott criterion (which is a position-momentum uncertainty argument), with $\hbar/\tau < k_\mathrm{B} T$, with the philosophy being that at finite temperature $k_\mathrm{B} T$ is the only dissipative energy scale.  Even after accepting this somewhat ill-defined uncertainty argument, it is unclear why the dimensionless coupling constant $\alpha$, which should be sitting in front of $k_\mathrm{B} T$ in such dimensional reasoning, should be of order unity since no theory constrains it and in principle, it could be anything.  But the two theoretical examples where detailed transport calculations are possible, namely the well-understood linear-in-$T$  metallic resistivity due to acoustic phonon scattering \cite{ziman1972principles, hwangLinearinResistivityDilute2019, wuPhononinducedGiantLinearin2019, dassarmaStrangeMetallicityMoir2022}\cmmnt{[ Ziman book, hwang-Das Sarma linear in T PRB 2019;   arXiv:1811.04920 ;   arXiv:2201.10270]} and low-temperature approximately linear-in-$T$ resistivity in 2D doped semiconductors (as well as the inelastic scattering in an electron liquid) in the current work, seem to obey the Planckian hypothesis unreasonably well (within a factor of $10$, $\alpha<10$), indicating that the effective coupling constant $\alpha$ entering the Planckian bound multiplying the temperature (in a strictly qualitative dimensional analysis) is indeed of order unity. Why it is so remains a mystery, and the possibility that this is merely a coincidence and that future experiments will discover strong violations of the Planckian hypothesis should not be ruled out. 
Our work suggests that looking for strong violations of the Planckian bound in any materials, by much more than an order of magnitude, should be a serious goal of future experiments so that we know for sure whether the generalized bound within an order of magnitude is generically valid or not.

\section{Acknowledgement} \label{sec:acknowledgement}
This work is supported by the Laboratory for Physical Sciences.
\clearpage


%

\end{document}